\documentclass[12pt,prd,aps,amssymb,amsmath,tightenlines,showpacs]{revtex4}
\usepackage{appendix,slashed}
\usepackage{graphicx,epsfig}
\usepackage{comment}
\usepackage{color}

\newcommand{\be}{\begin{equation}}
\newcommand{\ee}{\end{equation}}
\newcommand{\bea}{\begin{eqnarray}}
\newcommand{\eea}{\end{eqnarray}}

\begin{document}
\graphicspath{{FIGURE/}}
\topmargin=0.0cm

\title{The lifetime of the electroweak vacuum and 
sensitivity to Planck scale physics}

\author{Vincenzo Branchina$^a$}\email{branchina@ct.infn.it}
\author{Emanuele Messina$^a$}\email{emanuele.messina@ct.infn.it}
\author{Marc Sher$^b$}\email{mtsher@wm.edu}

\affiliation{${}^a$Department of Physics, University of Catania 
and INFN, Sezione di Catania, Via Santa Sofia 64, I-95123 
Catania, Italy\\
${}^b$High Energy Theory Group, Department of Physics, College of William and Mary, Williamsburg, VA  23187-8795 }

\begin{abstract}

If the Standard Model (SM) is valid up to extremely high energy scales, then the Higgs potential becomes unstable at approximately $10^{11}$ GeV.  However, calculations of the lifetime of the SM vacuum have shown that it vastly exceeds the age of the Universe.   It was pointed out by two of us (VB,EM) that these calculations are extremely sensitive to effects from Planck scale higher-dimensional operators and, without knowledge of these operators, firm conclusions about the lifetime of the SM vacuum cannot be drawn.   The previous paper used analytical approximations to the potential and, except for Higgs contributions, ignored loop corrections to the bounce action.   In this work, we do not rely on any analytical approximations and consider all contributions to the bounce action, confirming the earlier result.   It is surprising that the Planck scale operators can have such a large effect when the instability is at $10^{11}$ GeV.    There are two reasons for the size of this effect.   In typical tunneling calculations, the value of the field at the center of the critical bubble is much larger than the point of the instability; in the SM case, this turns out to be numerically within an order of magnitude of the Planck scale.  In addition, tunneling is an inherently non-perturbative phenomenon, and may not be as strongly suppressed by inverse powers of the Planck scale.   We include effective $\Phi^6$ and $\Phi^8$ Planck-scale operators and show that they can have an enormous effect on the tunneling rate.
\end{abstract}

\date{\today}

\maketitle   

\vfill
\eject

\section{Introduction}
\label{intro}

Shortly after the Standard Model (SM) was established, it was pointed out in a seminal paper by Cabibbo et al.\,\cite{cabibbo} that the quartic scalar coupling could either become non-perturbative or become negative before the unification scale is reached.  In the former case, new physics would have to intervene, and in the latter case the potential would become metastable; requiring that neither of these occur led to bounds on the Higgs and fermion masses.   Over the decades, this calculation has been increasingly refined \cite{flores,lindner,bennet,
sherrep,lindsher,arnold,anderson,arnoldvok,ford,sher2,altar,
casas1,espiquiros,casas2,frogniel1,isido,frogniel2}.   

While several different scenarios for physics beyond the Standard Model
are possible, the conservative choice is to assume that the Standard Model 
is  valid all the way up to the Planck scale $M_P$, i.e. 
that new physics interactions only occur at $M_P$.  This 
has been most recently investigated in Refs. 
\cite{espigiu,ellisespi,isiuno,isidue,degrassi}. 
According to these analyses, the recently measured value of 
the Higgs boson mass\,\cite{atlas,cms} is, in conjunction with 
improved measurements of the top quark mass, tantalizing close 
to the stability/metastability boundary. These calculations, 
however, show that the instability does occur at scales below 
the Planck scale.

The instability is primarily due to the top quark mass.    
Due to the loop corrections coming from the top, the Higgs 
effective potential $V_{eff}(\phi)$ turns over for values 
of $\phi$ much larger than $v$, the location of the 
electroweak (EW) minimum, 
and develops a new minimum at $\phi_{min} >> v$.
Depending on SM parameters, in particular 
on the top and Higgs masses, $M_t$ and $M_H$, the second 
minimum can be higher or lower than the EW one. In the 
first case, the EW vacuum is stable, in the second one it 
is metastable and we have to consider its lifetime $\tau$.
Normalizing $V_{eff}(\phi)$ so that it vanishes at $\phi=v$, 
in the case when $V_{eff}(\phi_{min}) < V_{eff}(v)$, the 
instability scale $\phi_{inst}$ is the value of $\phi$ 
such that $V_{eff}(\phi_{inst})=0$: for $\phi > \phi_{inst}$,  
the potential becomes negative, later developing the new 
minimum. For the Higgs and top masses 
given by the current central experimental values, 
$M_H \sim 125.7$ GeV and $M_t \sim 173.34$ GeV, 
$\phi_{inst} \sim 10^{11}\, {\rm GeV} >> v $. 

\begin{figure}[t]
\vskip-4mm
\includegraphics[width=.64\textwidth]{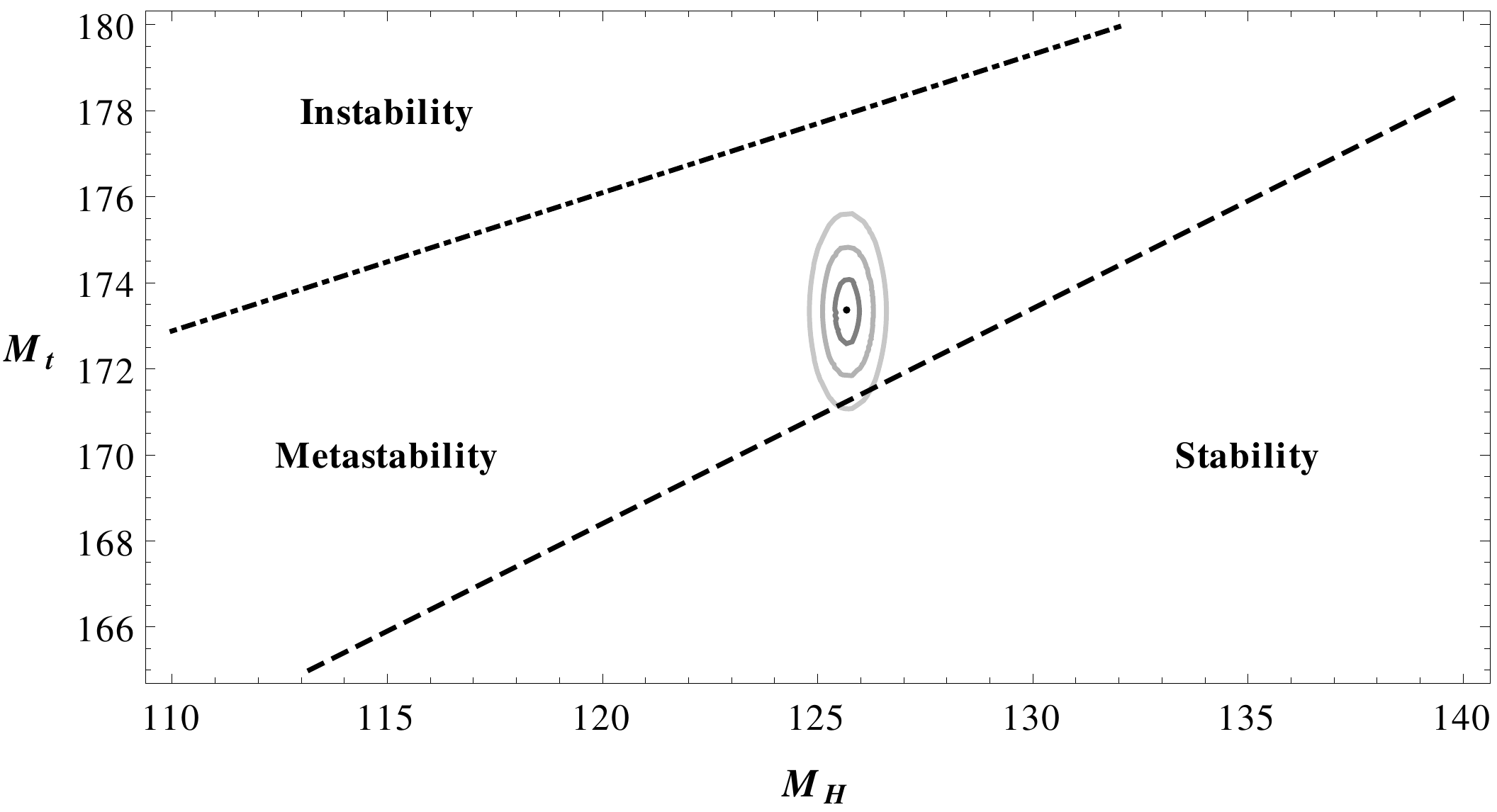}\vskip-4mm
\caption{
In this figure we plot the stability phase diagram 
according to the standard analysis, i.e. in the  
absence of new interactions at the Planck scale. The 
$M_H-M_t$ plane is divided in three sectors: absolute 
stability, metastability and instability regions. 
The dot indicates $M_H\sim 125.7$ GeV 
and $M_t\sim 173.34$ GeV. The ellipses take into account 
$1 \sigma$, $2 \sigma$ and $3 \sigma$, according to the 
current experimental errors. 
\label{bounn2}}
\end{figure}

The results are usually summarized with the help of the 
stability phase diagram of fig.\ref{bounn2}, where the 
$(M_H,M_t)$-plane is divided into three different sectors: 
an absolute stability region, where 
$V_{eff}(\phi_{min}) > V_{eff}(v)$, a (so called)
metastability region, where $V_{eff}(\phi_{min}) < V_{eff}(v)$, 
but the lifetime, $\tau$, is given by $\tau > T_U $, 
and an instability region, where 
$V_{eff}(\phi_{min}) < V_{eff}(v)$ but   
$\tau < T_U $ ($T_U$ is the age of the universe). 
The stability (dashed) line separates the stability  and 
the metastability sectors.  The instability (dotted-dashed) line
separates the metastability and the instability
regions and is 
obtained for $M_H$ and $M_t$ such that $\tau = T_U $.

This stability phase diagram is obtained by considering SM 
interactions only, as it is usually argued 
\cite{espigiu,ellisespi,isido,isiuno,isidue} 
that new physics interactions at 
the Planck scale, although present, have no impact on it.    
This argument seems quite reasonable, since the instability 
occurs at scales of $\sim 10^{11}\ {\rm GeV}$ and new 
physics interactions are suppressed by powers of the 
inverse Planck scale.
If this is really the case, from fig.\ref{bounn2},  we learn 
that for the current experimental values of $M_H$ and $M_t$, 
the electroweak vacuum is metastable, with a lifetime much larger 
than the age of the universe \cite{isido,isiuno,isidue}, and also 
that we are very close to the stability line (so called 
``criticality''), so that a better determination of
$M_H$ and $M_t$ would allow us to discriminate between a 
metastable, a stable or a critical vacuum state for 
our universe \cite{abdel,degrassi2}. 
Some authors consider this ``near criticality'' of the SM 
as the most important message from the data on the Higgs 
boson\,\cite{degrassi}. We note that this is also needed 
for the Higgs inflation scenario of \cite{ber}. 

For $M_H = 125.7$ GeV and $M_t \sim 173.34$ GeV, 
$\phi_{inst} \sim 10^{11}$ GeV. For $\phi>\phi_{inst}$, 
$V_{eff}(\phi)$ is negative and decreasing. For  
$\phi \geq M_P$,  the potential continues to decrease 
for a long while, forming a new minimum 
at a scale $\phi_{min}$ much larger than $M_P$, 
$\phi_{min} \sim 10^{30}\, {\rm GeV}$.   Of course, one expects Planck scale operators to have an effect long before that scale is reached.

It is usually argued \cite{isido} that this potential 
must be eventually stabilized by the unknown new physics around 
$M_P$. In other words, these new physics interactions are expected 
to modify $V_{eff}(\phi)$ around $M_P$ in such a way as to lead to a  new minimum around this scale. However, it is also
argued that the computation of the lifetime $\tau$ of the electroweak vacuum 
can still be performed with the help of the unmodified Higgs potential 
$V_{eff}(\phi)$, obtained with SM interactions only.  

As the instability occurs for very large values of $\phi$ 
($\phi_{inst} \sim 10^{11}$ GeV), $V_{eff}(\phi)$ is well 
approximated by keeping only the quartic term\,\cite{sher2}. 
Therefore, following\,\cite{cole1,Frampton:1976kf, cole2}, 
the electroweak vacuum lifetime 
is computed by considering first the bounce solution to the 
euclidean equation of motion for the classical potential $V(\phi)= 
\frac{\lambda}{4}\phi^4$ with a negative value of $\lambda$, and 
then taking into account the quantum fluctuations around the 
bounce. 

It has been recently shown, however, that new physics 
at $M_P$ can enormously modify the tunneling time and, more 
generally, the stability phase diagram \cite{brames, bra, bramespla}.
For the purposes of illustrating this effect, 
the analysis in\,\cite{brames}  was performed by considering   
two major simplifications. An 
approximation for the modified Higgs potential was considered 
that allowed for the existence of analytical bounce solutions;
and only the   
quantum fluctuations coming from the Higgs sector were considered.    

In the present paper, the analysis of\,\cite{brames} 
is improved, extended and completed in the following important 
aspects. First of all, we do not consider any approximation for 
the potential. Therefore, as we can no longer rely on analytical 
tools, we look for numerical bounce solutions for the complete 
potential. 
Also, the quantum fluctuation corrections to $\tau$
are computed by considering the contributions from all of the 
different sectors of the theory. 
This more complete analysis, as we shall see, confirms the 
results presented in\,\cite{brames} and provides 
the theoretical support for the results presented in 
\cite{bramespla}, where some of the results presented in this 
work were anticipated and used. 

The rest of the paper is organized as follows.  In the next 
section, we review the calculation of the electroweak vacuum
lifetime in the Standard Model. It is shown there that 
the standard assumption 
that Planck scale operators can be neglected may not be valid, 
since the value of the field in the center of the critical 
bubble is much larger than the instability scale, and is 
close to the Planck scale.  In section III, the effects of 
Planck scale operators are then included.  In section IV, 
we compare the numerical results with the analytic results 
of Ref. \cite{brames}, and section V contains our conclusions.   
There are three appendices. In Appendix A, the computation of 
the quantum fluctuation contribution to the tunneling time 
is presented in some detail. Appendix B provides some 
tools for the numerical computation of the bounce. In 
particular, the bounce considered in section III
is computed. In Appendix D, we provide a explicit example, 
using $SU(5)$, giving the size of the higher dimensional 
operators.

\section{Bounces and the Planck scale $M_P$}
\label{known}

Before starting our analysis on the impact of new physics, 
in the present section we focus our attention on the 
standard analysis, where it is assumed that the stability phase 
diagram and, in particular, the lifetime of the electroweak vacuum  $\tau$
are not affected by new physics at the Planck scale 
\cite{espigiu,ellisespi,isido,isiuno,isidue}. 

Let us begin by considering the euclidean action for the 
scalar sector of the SM

\begin{equation}
S[\Phi]=\int\,d^4x\,
(\left(\partial_\mu\Phi\right)^\dagger \cdot \left(\partial_\mu\Phi\right) 
+V(\Phi) )\label{actionSec}
\end{equation}
where we write the scalar doublet $\Phi$ as
\begin{equation}\label{doublet}
\Phi=
\frac{1}{\sqrt{2}}\left(\begin{array}{c}
-i(G_1-iG_2)\\
\phi+i G_3
\end{array}\right),
\end{equation}
with $\phi$ the Higgs field and $G_i$ the Goldstone bosons, while the 
potential 
$V(\Phi)$ is, for large values of $\phi$,
\begin{equation}\label{appropote}
V(\Phi)=\lambda (\Phi^\dagger\Phi)^2\,.
\end{equation}

The procedure for determining the tunneling rate was first discussed in 
Refs. \cite{cole1,Frampton:1976kf,cole2}, and a very clear discussion involving the Standard Model can be found in Ref. \cite{Kusenko:1996bv}. 
The bounce, $\phi_b$, is a solution of the Euclidean equations of motion 
for the above action.
Renaming for a moment $S$ as the full SM action, 
following\,\cite{Kusenko:1996bv} we write for the tunneling 
probability (details are given in Appendix A)

\begin{eqnarray}\label{tunnel}
p = \int {\prod_{i=1}^8} \,d \gamma_i \,\, 
J_{zeros}(\gamma_1,...,\gamma_8)
\left|\frac{SDet'(S''[\phi_b])}{SDet(S''[0])}\right|^{-1/2}
e^{-S[\phi_b]}\,.
\end{eqnarray}

$S[\phi_b]$ is the tree-level action computed at $\phi=\phi_b$,
with all of the other SM fields vanishing. $S''$ denotes double 
functional differentiation with respect to all of the  
SM fields. $SDet$ is the Superdeterminant, and $Det^\prime$ means 
that in the computation of the determinant the zero modes are 
excluded ($SDet(S''[0])$ comes from the normalization). The 
$\gamma_i$ ($i=1,...,8$) are the collective coordinates, 
the flat directions related to the zero modes, and  
$J_{zeros}(\gamma_1,...,\gamma_8)$ is the product of the 
Jacobians coming from 
the corresponding change of variables in the path integral 
(from usual to collective coordinates). In the SM there are 
eight zero modes: four translational (the collective 
coordinates being $x_0$, $y_0$, $z_0$, $t_0$, the coordinates
of the center of the bounce), three related to $SU(2)$
``rotations'' (the collective coordinates being the angles
$\theta_1$,$\theta_2$ and $\theta_3$) and finally, when 
the potential is taken as in Eq.\,(\ref{appropote}) (where 
the mass term is neglected), one dilatation zero mode (the
collective coordinate being the size $R$ of the bounce).    
The complicated term in front of the exponential is often 
sub-dominant, although we will include it here.

\begin{figure}[t]
\vskip-4mm
\includegraphics[width=.64\textwidth]{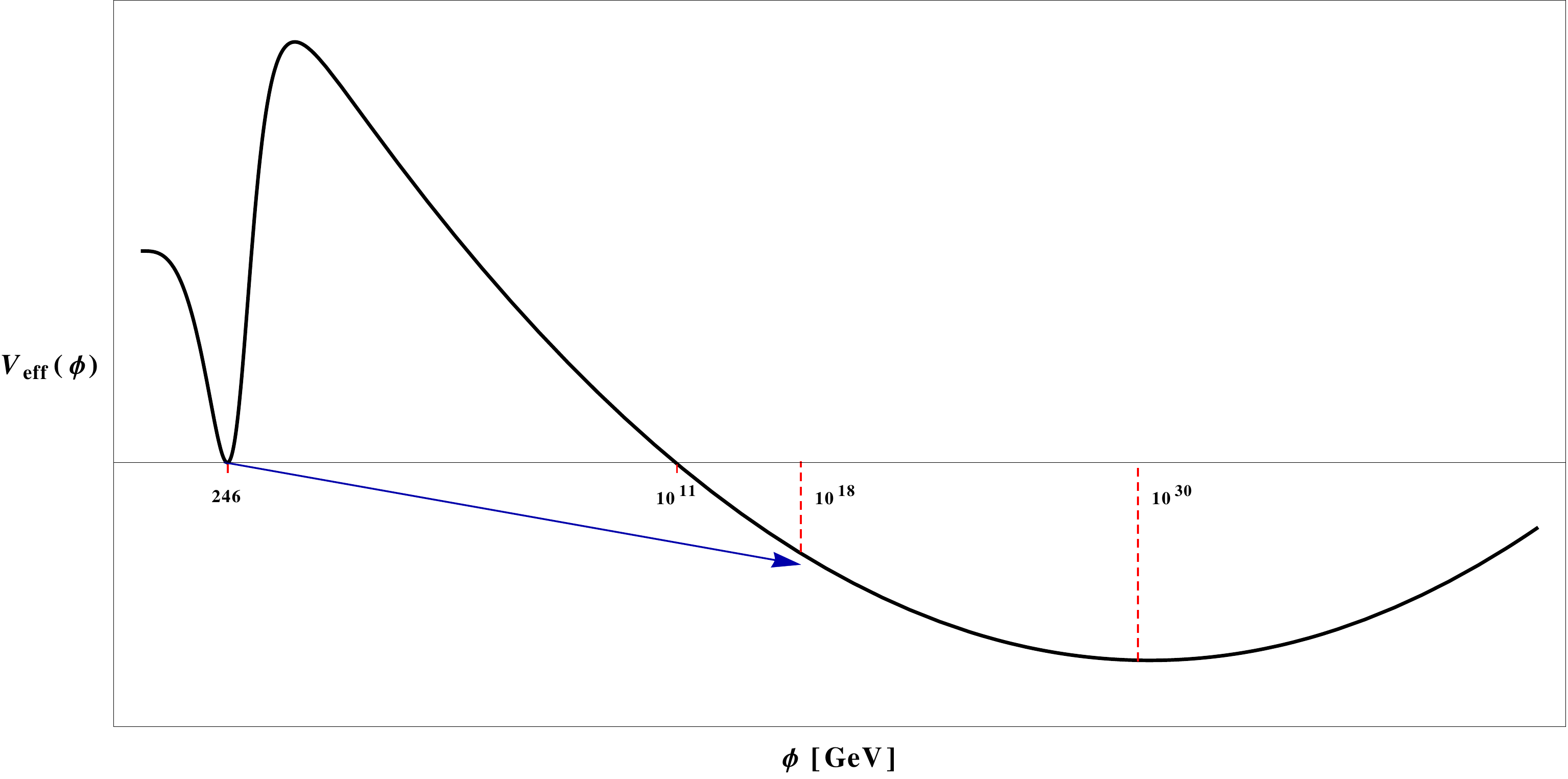}\vskip-4mm
\caption{The potential in the Standard Model, for $M_H =125.7$ GeV 
and $M_t=173.34$ GeV, is sketched (figure not to scale). 
The potential goes negative at a scale of $10^{11}$ GeV and 
reaches a new minimum at roughly $10^{30}$ GeV. The tunneling 
through the barrier goes from the base of the arrow 
($\phi(r=\infty)$) to the tip ($\phi(0)$), which turns out to 
be close to or above the Planck scale.
}\label{bounn}
\end{figure}

For negative values of $\lambda$, the (euclidean) equation 
of motion for the action (\ref{actionSec}) has non-trivial
configuration solutions for the Higgs field (with $G_i=0$), 
i.e. bounce solutions, which are 
solutions of the equation ($r$ is the radial coordinate 
in ${\mathbb{R}}^4$)
\begin{equation}
\frac{d^2\phi}{dr^2}+\frac{3}{r}\frac{d\phi}{dr}-\frac{dV}{d\phi}=0\,, 
\label{eqbounce}
\end{equation}
with boundary conditions
\begin{equation} 
\phi(\infty)=0
\end{equation}
\begin{equation} 
\left.\frac{d\phi(r)}{dr}\right|_{r=0}=0 \,,
\end{equation}
where $V(\phi)$ is
\begin{eqnarray}
V(\phi)=\frac{\lambda}{4}\phi^4\,. \label{pot}
\end{eqnarray}

Note that Eq. (\ref{eqbounce}) is also obtained by considering 
the restriction
\begin{equation}
S[\phi]=\int\,d^4x\,(\frac12\partial_\mu\phi\partial_\mu\phi 
+V(\phi) )\label{action}
\end{equation}
of the action (\ref{actionSec}) when all the $G_i$ vanish. 

The family of bounce solutions to Eq.\,(\ref{eqbounce}) is
\begin{equation}
\phi_b(r)=\sqrt{\frac{8}{|\lambda|}}\frac{R}{r^2+R^2}\,,\label{bouncecla}
\end{equation}
and is parametrized by $R$, the size of the bounce ($0<R<\infty$).

For negative values of $\lambda$, the action (\ref{action}) is scale
invariant, so that all these 
configurations, irrespectively of the size $R$, have the same value 
of the action, namely
\begin{equation}
S[\phi_b]=\frac{8\pi^2}{3|\lambda|} \,.\label{actscaleinv}
\end{equation}
From Eq.\,(\ref{bouncecla}), we see that $R$  
and $\phi_b(0)$ (the maximal value of $\phi_b(r)$) 
are related by 
\begin{equation}\label{err}
R=\sqrt{\frac{8}{|\lambda|}}\frac{1}{\phi_b(0)}
\end{equation}
and that $R$ is nothing but that value of $r$ such that 
\begin{equation}\label{size}
\phi_b(R)=\frac12\,\phi_b(0).
\end{equation}

In Figure \ref{bounn}, we have sketched the potential. Note that the 
tunneling does not lead directly to the other side of the 
barrier.  This is because of the gradient terms (surface 
tension for a thin-walled bubble), which require the bubble 
to gain volume energy. The point at the tip of the arrow 
is $\phi_b(0)$. The value of $\phi_b(0)$ can, in principle, 
be substantially larger than the point of the instability, 
and we will shortly see that this does, in fact, occur.

Going back to Eq.\,(\ref{tunnel}), we note that the integration 
over the center of the bounce (the four translational
zero modes) can be immediately performed and gives the four-volume
factor $\Omega= V\,T_U$ ($V$ and $T_U$ are the volume and the age 
of the universe, respectively), that in 
our case is  $\Omega= T_U^4$. The same is true for the integration
over the angular $SU(2)$ variables ($\theta_1$, $\theta_2$, 
$\theta_3$), that provides a factor $16\pi^2$. 

Finally, concerning the integration in the remaining collective 
coordinate, the bounce size $R$, we note that, although the 
value of $S$ is the same for all bounce sizes, $R$, quantum 
fluctuations break the degeneracy, and only one value 
of $R$, say $R_M$, saturates the path integral. 

Therefore, from Eq.\,(\ref{tunnel}) for the tunneling probability, 
we can immediately write the tunneling time as
\be\label{life}
\tau=\left[
\frac{R_{M}^4}{T_U^4}\,e^{\,\frac{8\pi^2}
{3|\lambda(\mu)|}}\right]\times 
\left[e^{\Delta S}\right] \times T_U \,,
\ee 
where we have used Eq.\,(\ref{actscaleinv}) for $S[\phi_b]$,  
and $\Delta S$ corresponds to quantum fluctuations, 
to be discussed shortly, 
\begin{eqnarray}\label{deltaes}
\Delta S = - ln \left(\frac{16\pi^2}{R^8} J_{trans} J_{SU(2)} 
J_{dil} \left|\frac{SDet'(S''(\phi_b))}{SDet(S''(0))}
\right|^{-1/2}\right)_{R=R_M}\,,
\end{eqnarray}
the Jacobian factor of Eq.\,(\ref{tunnel}) being split
into the product of the three Jacobians related to the 
translation, dilatation, and $SU(2)$ zero modes 
(in Appendix \ref{App2} these Jacobian factors,  
together with the determinants, are computed).  

Crucial to our analysis is the knowledge of the running of the 
quartic coupling $\lambda(\mu)$, to be solved together with the 
coupled RG equations 
for the other SM couplings. We have used the RG equations 
up to the next-to-next to leading order. The beta functions 
and the boundary conditions up to this 
order have been recently worked out and are presented  
in \cite{miha, chety, shapo, isidue}. 

By considering the RG equations for $\lambda(\mu)$, 
we see that the instability of the kind 
shown in Figure 2 occurs when
$\lambda(\mu)$ hits zero and then becomes 
negative. This is the case when the electroweak vacuum is 
metastable. For sufficiently large
values of $\mu$, $\lambda(\mu)$ saturates to a constant 
negative value. As for the renormalization scale 
$\mu_{ren}$, it is convenient to choose $\mu_{ren} \sim 1/R_M$. 
This is the value of $\lambda(\mu)$ to be used 
in Eq.\,(\ref{life}).   
For $M_H=127.5$ GeV and $M_t=173.34$ GeV, we find 
\be\label{rmax} 
R_M\sim 1.87 \cdot 10^{-17}\,GeV^{-1} = 224.5\,M_P^{-1} 
\ee 
and 
\be\label{lambdar}
\lambda (1/R_M) = -0.01345\,, 
\ee
that in turn gives 

\begin{equation}
S[\phi_b]= 1956.54\,. \label{actscaleinvnum}
\end{equation}

Inserting Eqs.\,(\ref{rmax}) and (\ref{actscaleinvnum}) in
Eq.\,(\ref{life}), 
a first estimate of $\tau$ can be 
obtained by considering 
the classical (tree level) contributions only, i.e. by 
neglecting the quantum fluctuations (the term 
$e^{\Delta S}$). We find that
\begin{equation}
\tau_{tree} \sim  10^{\,613}\,T_U\,. \label{lifeclass}
\end{equation}

At tree level, we already see that the electroweak vacuum lifetime 
$\tau $ turns out to be enormously larger than the age of the 
universe, thus justifying the so called metastability scenario:
the electroweak vacuum is metastable but its lifetime is much larger than
the age of the universe. This is why the allowed region in 
Figure \ref{bounn2} is so far from the line where the lifetime is the 
age of the Universe.

The next step is the inclusion of the quantum fluctuations.
In Eq.\,(\ref{life}), the contribution of the  
fluctuation determinant is given by the factor 
$e^{\Delta S}$. More precisely, each of the different sectors 
of the theory (Higgs, gauge, goldstone, top) provides a 
contribution to $\Delta S$, which then takes the form
\begin{equation}\label{deltaS}
\Delta S= \Delta S_H + \Delta S_t + \Delta S_{gg}\,,
\end{equation}
where $\Delta S_H$ is the loop contribution from he Higgs sector,
$\Delta S_t$ the contribution from the top sector and  
$\Delta S_{gg}$ the one from the gauge and Goldstone sectors. 

\begin{figure}
${}$\vskip1cm \epsfig{file=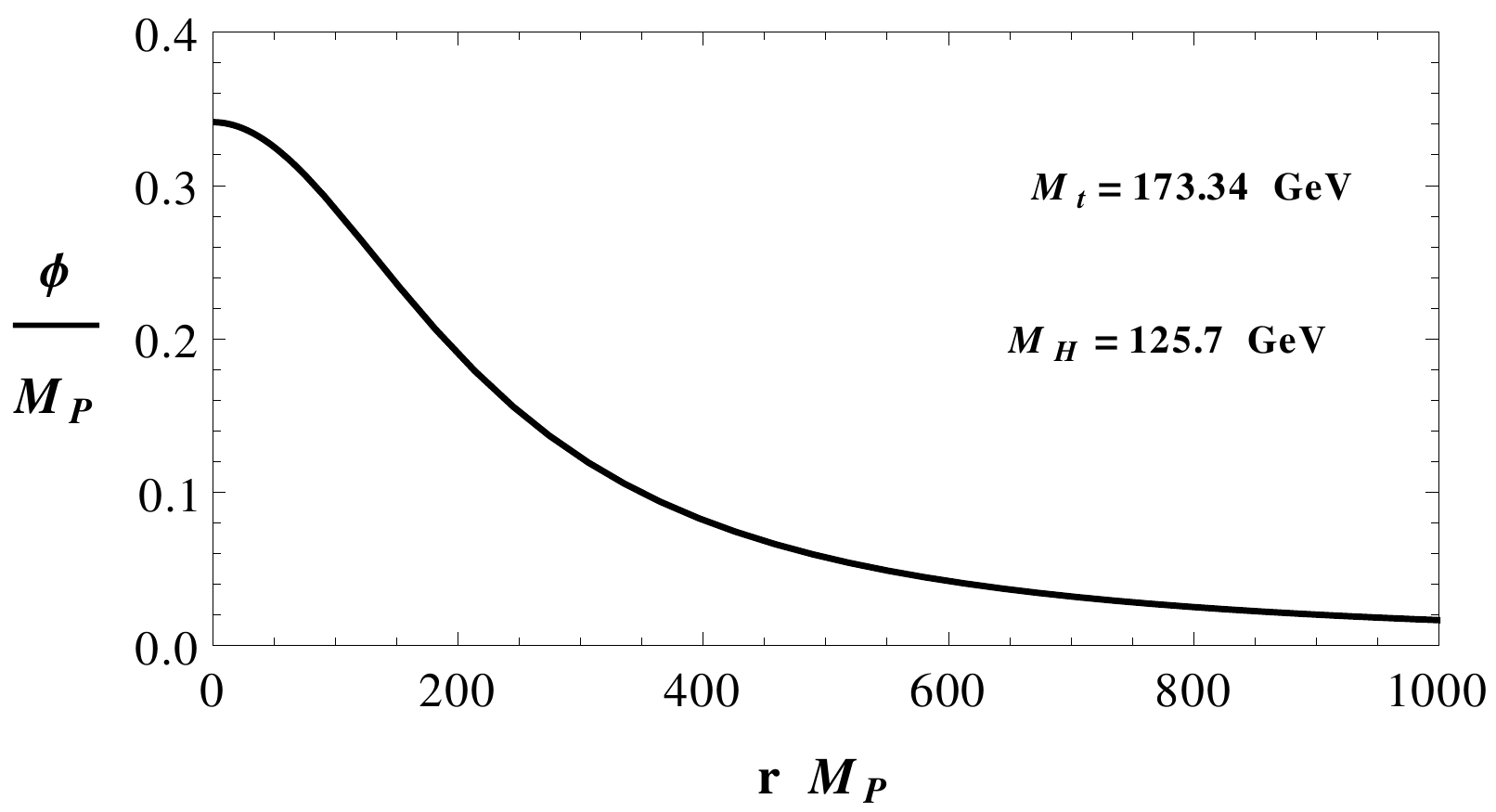,width=.75\hsize}
\caption{Profile of the bounce solution that enters in the 
computation of the electroweak vacuum lifetime $\tau$ for 
$M_H= 125.7$ GeV and $M_t=173.34$ GeV, the present central 
experimental values of $M_H$ and $M_t$. The value of the 
field at the center of the bounce
($r=0$) is $\phi_b(0)=0.34\, M_P$, very close to the Planck 
scale. }
\label{bounc}
\end{figure}

In Appendix \ref{App2} the computation of the different 
$\Delta S_i$ is shown. Here we present the results in the table 
below   

\begin{center}
\begin{tabular}{cc}\label{tab1}
 & Loop contributions to $\tau$ \\
\hline
$e^{\Delta S_H}~$ & $10^{-7}$  \\
$e^{\Delta S_t}~$ & $10^{-19}$ \\
$e^{\Delta S_{gg}}$ & $10^{68}$ \\
\end{tabular}
\end{center}

Collecting the different multiplicative 
contributions to $\tau$ listed above, we  finally have 
\begin{equation}
\tau \sim  10^{655}\,T_U\,. \label{lifequant}
\end{equation}

Despite of the enormous difference in magnitudes between 
(\ref{lifeclass}) and (\ref{lifequant}), it seems appropriate 
to quantify the distance between the classical and the 
quantum corrected estimate of $\tau$ by noting that in terms of 
orders of magnitudes, the exponent $655$ in (\ref{lifequant})
provides a 6 $\%$ correction to the exponent $613$ in 
(\ref{lifeclass}). In this sense, even the 
tree level result (\ref{lifeclass}) gives, in this framework, 
a ``good'' estimate of $\tau$. 

What we have just seen is that, even after the inclusion of the 
quantum fluctuation corrections, the lifetime of the electroweak vacuum
$\tau $ turns out to be enormously larger than the age of the 
universe, and this seems to give support to 
the metastability scenario.   As explained in the introduction, a 
more complete study of electroweak vacuum stability can be done  
in terms of the Higgs and top masses $M_H$ and $M_t$. In 
Figure \ref{bounn2}, the corresponding SM phase diagram  in the 
$M_H-M_t$ plane is shown. 

\begin{figure}[t]
$$\includegraphics[width=0.45\textwidth]{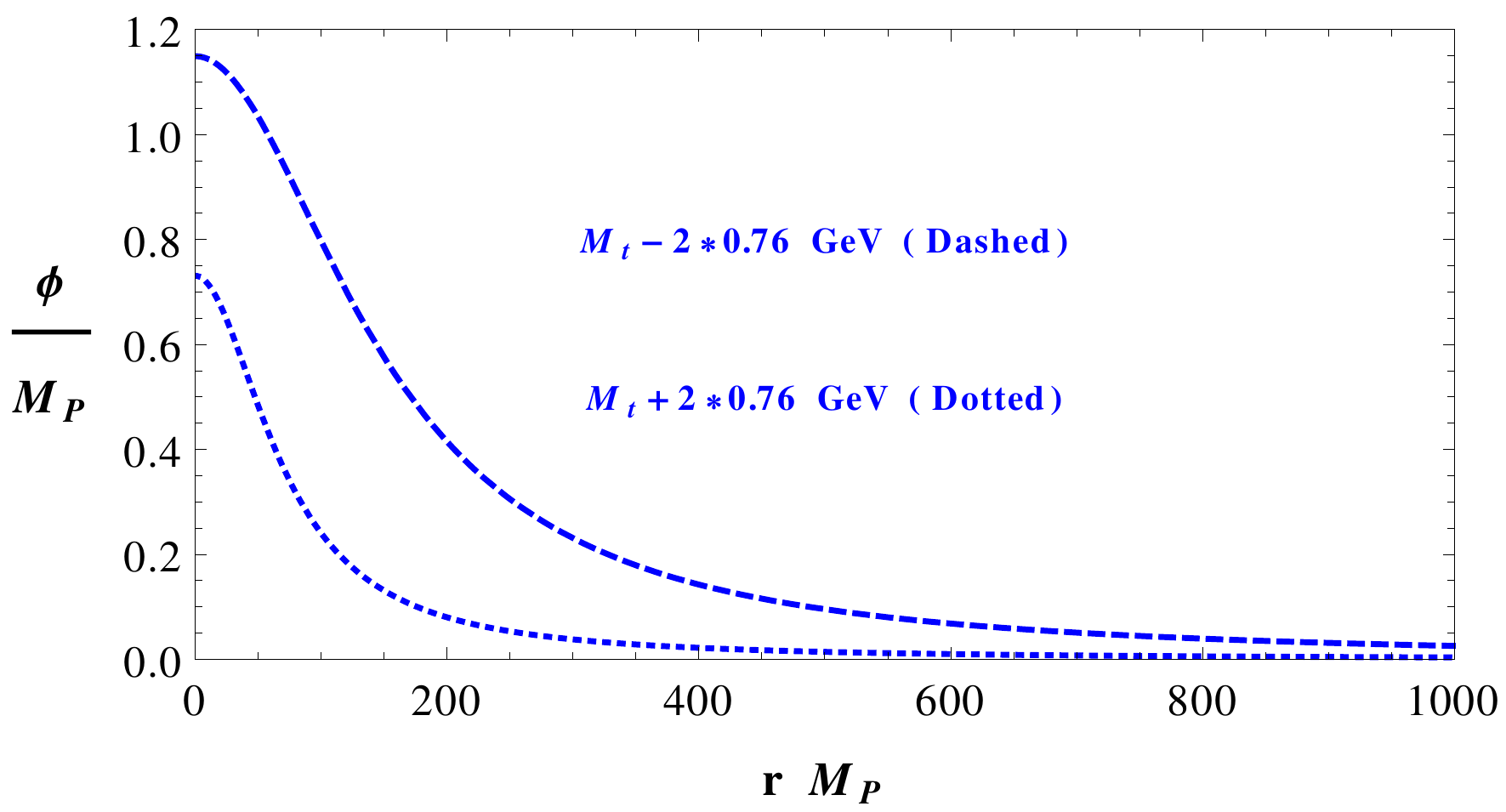}\qquad
\includegraphics[width=0.45\textwidth]{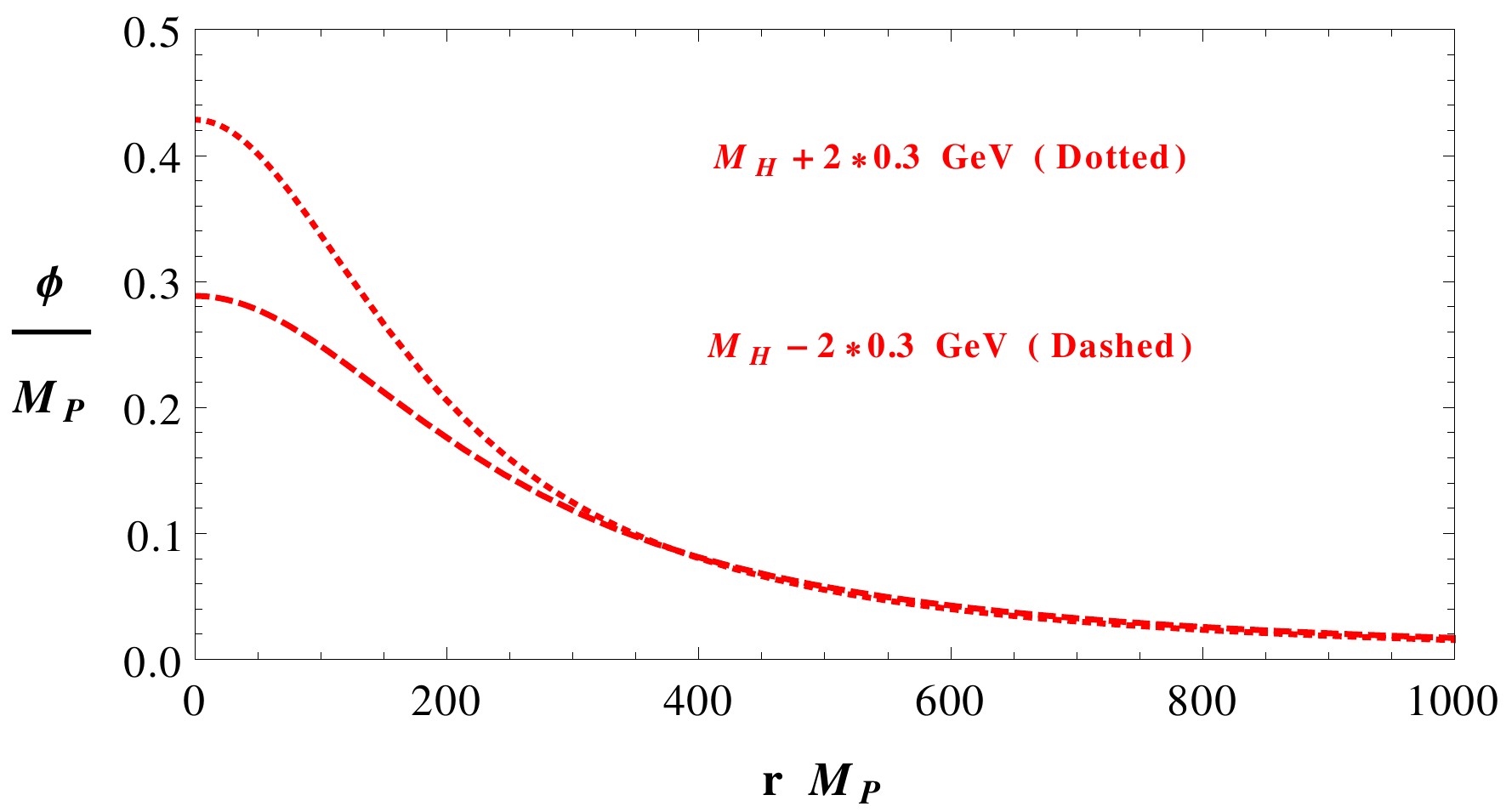}$$
\caption{Profile of the bounce solution that enters in the 
computation of the electroweak vacuum lifetime $\tau$ for
values of $M_H$ and $M_t$ slightly different from those 
of Figure \ref{bounc}. Actually, $\pm 2\sigma$ (current 
experimantal errors) for $M_t$ in the left
panel (with $M_H$ kept fixed to the central value 
$M_H=125.7$ GeV), and $\pm 2\sigma$ (current experimental errors) 
for $M_H$ in the right
panel (with $M_t$ kept fixed to the central value 
$M_t=173.34$ GeV). 
As in Figure \ref{bounc}, the values of the field at the center
of the bounce, $\phi_b(0)$, turn out to be very close to the 
Planck scale, sometimes even above this scale.  
\label{bouncealt} }
\end{figure}

We now move to consider one of the key points of this paper, 
by turning our attention to the profile of the bounce.
As we said above, due to the removal 
of the degeneracy from quantum fluctuations, 
the path integral for the computation of $\tau$ is saturated 
by only one of the bounces, with a specific value of the 
size $R$, $R_M$. For $M_H = 125.7$ GeV and $M_t = 173.34$ GeV, 
$R_M$ is given in Eq.\,(\ref{rmax}). Moreover, the value of the 
quartic coupling for the same values of $M_H$ and $M_t$ is given 
in Eq.\,(\ref{lambdar}). Then, from Eq.\,(\ref{bouncecla}),
we can determine the profile of the bounce that enters the 
evaluation of $\tau$.  The result is given in 
Figure \ref{bounc}. We have also shown the profiles for different 
values of $M_H$ and $M_t$ in Figure \ref{bouncealt}.  

Looking at these results, we see that the value of the field 
at the center of the bubble, $\phi_b(r=0)$, is dangerously 
close to the Planck scale. One can then suspect that Planck 
scale effects might be significant, even though the 
potential becomes unstable at a scale of roughly 
$10^{-8} M_P$, i.e. much below $M_P$. In this 
respect, it is important to note that the Planck mass never 
entered into our calculation, we have simply scaled $\phi$ 
and $r$ in terms of $M_P$, instead of $GeV$ and $GeV^{-1}$ 
respectively.   

The key point that emerges from inspecting these bounce 
profiles (figs. \ref{bounc} and \ref{bouncealt}), then,
is that the value of the field at 
the center of the bubble can be not only substantially larger 
than the instability scale, but actually so close to 
$M_P$ that Planck scale effects can be expected to
affect the tunneling rate. 
In order to investigate this question, we will now add Planck 
scale operators 
to the potential and redo the calculation.
We will see in the next section that 
the  results (\ref{lifeclass}) and (\ref{lifequant}) on the electroweak
vacuum lifetime and the phase diagram of Fig.\,\ref{bounn2}
can be dramatically modified.

\section{Bounces and new physics}
\label{higherpow}

In order to study the impact of new 
physics interactions at the Planck scale on the electroweak vacuum 
lifetime $\tau$, following \cite{brames,bra,bramespla}, 
we consider a simple modification of the 
theory by adding to 
the quartic potential (with negative $\lambda$) of the previous 
section two higher powers of the scalar field 
\begin{eqnarray}
V_{new}(\phi)=\frac{\lambda}{4}\phi^4
+\frac{\lambda_6}{6 M_P^2}\phi^6
+\frac{\lambda_8}{8 M_P^4}\phi^8\,.  \label{potnew}
\end{eqnarray}

The goal of the present work is not that of 
studying specific models. Our aim is rather to show that the 
presence of new physics
at the Planck scale is far from being harmless in the 
evaluation of the electroweak vacuum lifetime. The choice of the
potential (\ref{potnew}) is well suited for this purpose.   
As a demonstration of a model in which this potential arises 
as an effective field theory (without, to leading order in 
the couplings, $\phi^{10}$ or higher terms), 
in Appendix \ref{App4} we have given 
an example from a minimal SU(5) model, in which $M_P$ is 
replaced by the unification scale.  This shows that it is very 
easy to have $\lambda_6$ and $\lambda_8$ of $O(1)$.
In order to have a stable potential, $\lambda_8$ has to
be taken positive, while $\lambda_6$ can have both signs.
In the toy minimal SU(5) model that we look at
in the Appendix \ref{App4} this happens automatically.

In contrast with the previous section, with the potential 
(\ref{potnew}) we cannot find 
analytical solutions to the euclidean equation 
of motion (\ref{eqbounce}). 
Moreover, 
the scale invariance of the action (\ref{action}) is lost.
However, when $\phi<<M_P$ and the coupling constants $\lambda_6$ 
and $\lambda_8$ have natural $O(1)$ values, (\ref{potnew}) is 
well approximated by (\ref{pot}). Under these conditions, 
the new action is {\it almost} scale invariant and the 
configurations (\ref{bouncecla}) turn out to be good 
{\it approximate solutions} even for $V_{new}(\phi)$.     
Note that as long as we limit ourselves 
to consider bounces of ``large size'' (large with respect to 
$1/M_P$), even in the presence of the higher order operators 
$\phi^6$ and $\phi^8$, the configurations 
(\ref{bouncecla}) are (quasi-)solutions
to the euclidean equation of motion (a result to be expected).

In the computation of the tunneling time, then, these 
configuration have to be taken into account. Will will 
come back to this point at the end of this section. 
But for now, let us look for the existence of exact bounce 
solutions to the euclidean equation of motion (\ref{eqbounce})
with the potential (\ref{potnew}). 
Although we cannot rely on analytical tools, with the help of 
forward-backward shooting techniques\,\cite{dunne}, we can 
search for numerical solutions. 

For our purposes, it is useful to rescale the radial coordinate $r$ 
and the field $\phi$ by defining the dimensionless coordinate $x$ and 
the dimensionless field $\varphi$ in terms of Planck mass units 
\begin{equation}\label{resccc1}
x={M_P}\, r
\end{equation}
\begin{equation}\label{resccc2}
\varphi(r)=\frac{\phi(x)}{M_P}\,.
\end{equation}
Eq.\,(\ref{eqbounce}), with the potential (\ref{potnew}), then 
becomes 
\begin{eqnarray}
\frac{d^2\varphi}{dx^2}+\frac{3}{x}\frac{d\varphi}{dx}
-\lambda\varphi^3-\lambda_6\varphi^5
-\lambda_8\varphi^7=0\,, \label{eqcom}
\end{eqnarray}
while the boundary conditions are
\begin{equation}\label{c1}
\varphi(\infty)=0
\end{equation}
\begin{equation}\label{c2} 
\left.\frac{d\varphi(x)}{dx}\right|_{x=0}=0 \,.
\end{equation}

\begin{figure}
${}$\vskip1cm \epsfig{file=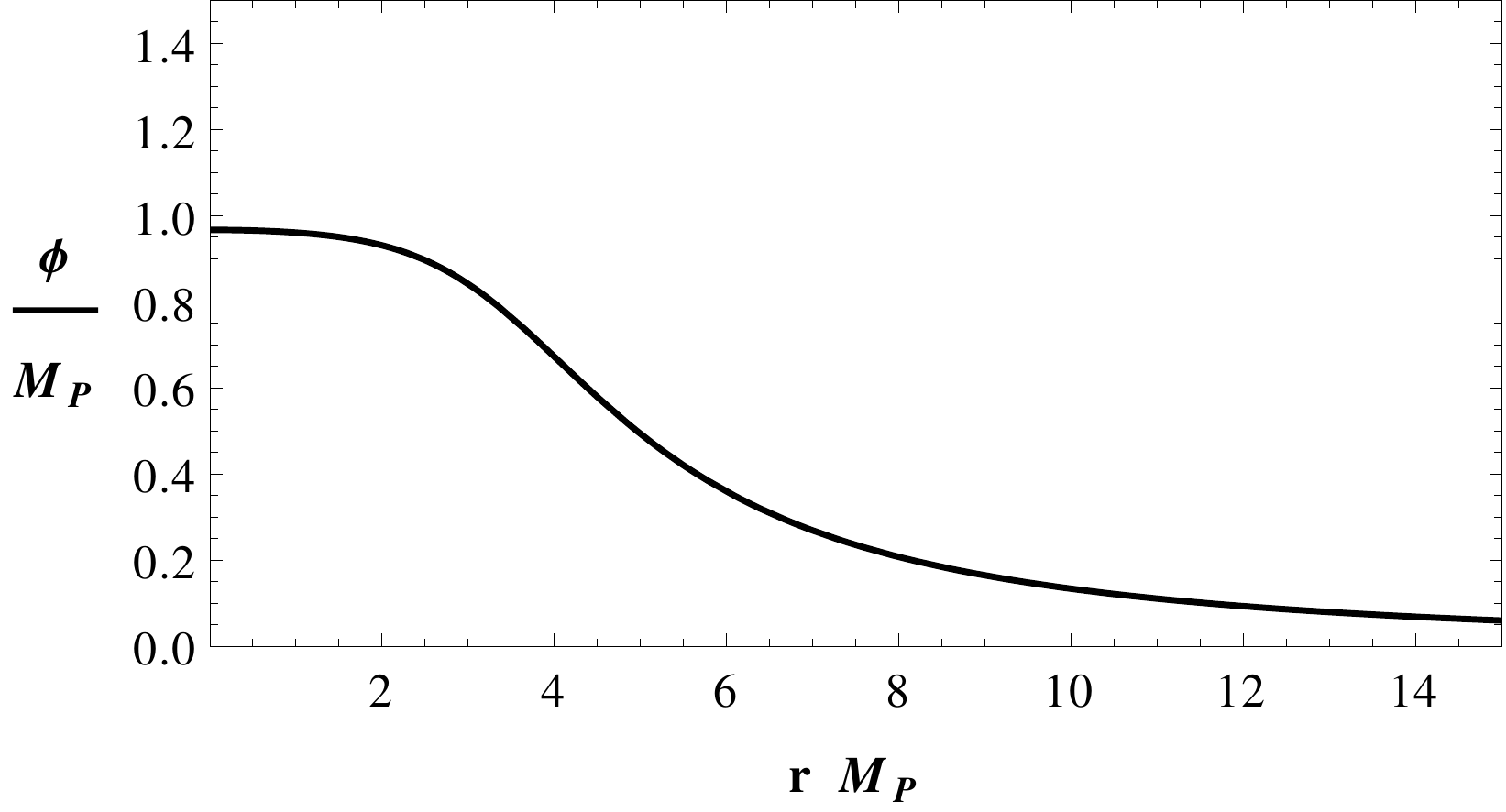,width=.75\hsize}
\caption{Profile of the bounce solution found with the 
forward-backward method described in Appendix \ref{App3}  
for the potential of Eq.(\ref{potnew}), with $\lambda=-0.01345$, 
$\lambda_6=-2$, and $\lambda_8=2.1$.}\label{bouncenoi}
\end{figure}

In Appendix \ref{App3}, 
Eq.\,(\ref{eqcom}) is solved numerically with the help of 
forward-backward shooting methods. 
The profile $\varphi_{_{bou}}(r)$ of the bounce solution found 
with the help of the numerical procedure outlined in this Appendix 
is plotted in Fig.\,\ref{bouncenoi}. 
Here we have somewhat arbitrarily chosen  $\lambda_6=-2$ and 
$\lambda_8=2.1$. This profile has to be compared with the bounce
of Fig.\,\ref{bounc}, which is a solution obtained for the 
potential (\ref{pot}), i.e. in the absence of the higher order 
operators $\phi^6$ and $\phi^8$. 
Quite interestingly, the value of the field at 
the center of the bounce, $\phi_b(r=0)$, is not much different 
from the values obtained for the case when the Planckian new
physics operators $\phi^6$ and $\phi^8$ are absent 
(see Figs. \ref{bounc} and \ref{bouncealt}).

Going back to dimensionful quantities, naming
$\phi_{_{bou}}(r)$ the dimensionful counterpart of 
$\varphi_{_{bou}}(r)$ (see (\ref{resccc1})) and  
defining the size $\overline R$ of this bounce 
according to (\ref{size}), i.e. as that value of $r$ such 
that 
\begin{equation}
\phi_{_{bou}}(\,\overline R\,)=\frac{1}{2}\phi_{_{bou}}(0)\,,\label{size2}
\end{equation}
we obtain  
\begin{equation}\label{size3}
\overline R \simeq 5.06\,M_P^{-1}\,.
\end{equation}
As for the corresponding action, from (\ref{action}) 
and (\ref{potnew}) we have 
\begin{eqnarray}
S[\phi_{_{bou}}]\simeq 82.09\label{actbounnum}\,.
\end{eqnarray}
Note that this action is much, much less than the action in Eq.(\ref{actscaleinvnum}), implying that the lifetime of the electroweak vacuum is much, much smaller.

Let us pause for a moment to make some comments. 
The classical theory considered in 
the previous section is scale invariant. This is why we found 
an infinity of bounce solutions with 
all possible values of the size. The quantum fluctuations  
lifted the degeneracy and the path integral was then dominated
by a single bounce with a well defined size $R_M$. In the present case, 
the classical theory with potential (\ref{potnew}) is no longer 
scale invariant. Accordingly, there is no degeneracy in the bounce 
size already at the classical level. Our numerical procedure, in 
fact, has shown that there is only one bounce,
with a well defined size $\overline R$, that solves the euclidean equation 
of motion and satisfies the boundary conditions 
for the bounce. This removal of the degeneracy at the classical 
level certainly  
occurs whenever new physics interactions at the Planck (or, more 
generally, new physics) scale are included.

Having at our disposal $\overline R$ and 
$S[\phi_{_{bou}}]$, we are in the position to compute, according 
to (\ref{life}), the tree-level contribution to $\tau$, i.e. 
the contribution obtained neglecting the quantum 
fluctuation ($\Delta S =0$)

\be\label{lifen}
\tau_{tree} \sim \left[
\frac{\overline R^4}{T_U^4}\,e^{S[\phi_{bou}]}\right]T_U 
\sim 10^{-206}\, T_U\,. 
\ee

Eq.\,(\ref{lifen}) is the key result. It has to be 
compared with Eq.\,(\ref{lifeclass}) of the previous section.
From this comparison we immediately see that the inclusion of
new physics interactions at the Planck scale, already at the 
classical (tree) level, has produced a dramatic modification 
in the electroweak vacuum lifetime. 
A bona fide computation where new physics interactions
at the Planck scale are explicitly taken into account has 
shown that they have a huge impact on 
the electroweak vacuum lifetime.   
Clearly, such values for $\lambda_6$ and $\lambda_8$ are 
phenomenologically unacceptable. This shows the importance 
of Planck scale operators on the metastability calculations, 
and shows that the conventional diagram of Fig. \ref{bounn2} 
can be drastically changed by such operators.

It might be surprising that the Planck scale operators can 
have such a large effect.  After all, while the value of the 
field at the center of the bubble is fairly close to the Planck 
scale, it isn't substantially larger (and most of the field 
values throughout the bubble wall are substantially smaller) 
and thus one might expect $O(1)$ corrections, not the huge 
corrections we have seen. However, one must keep in mind that 
tunneling is a non-perturbative phenomenon. The tunneling 
rate is computed by looking for the bounce solution and then 
considering quantum fluctuations on top of that.  While 
the latter are perturbative, and thus suppressed by inverse 
powers of the Planck scale, the former is not.   

The potential (\ref{potnew}) differs from the potential
$\lambda\phi^4/4$, and the corresponding new saddle point 
$\phi_{bou}$ provides a different non-perturbative 
contribution $e^{-S[\phi_{bou}]}$ to the tunneling 
rate. The bounce $\phi_{bou}(r)$ is a profile, not a localized 
configuration, defined in the whole range $r\in[0, \infty[$.
No matter how similar it looks to $\phi_b(r)$ of the previous 
section. The difference between these two profiles provides the 
difference between the two exponentials $e^{-S[\phi_{b}]}$
(previous section) and $e^{-S[\phi_{bou}]}$ (this section),
and these two numbers are exponentially decoupled. 

As in the previous section, the next step consists  
in the inclusion of the quantum fluctuations. Once again, the 
contribution of the fluctuation determinant is given in terms 
of the factor $e^{\Delta S}$ and, as before, each of the different 
sectors of the theory (Higgs, gauge, goldstone, top) provides a 
contribution to $\Delta S$ 
($\Delta S= \Delta S_H + \Delta S_t + \Delta S_{gg}$).
These are computed in Appendix \ref{App2}.
Here we present the results in the table below     

\begin{center}
\begin{tabular}{ccc}
 & Loop contributions to $\tau$ \\
\hline
$e^{\Delta S_H}~$   & $ 10^{-9}$  \\
$e^{\Delta S_t}~$   & $ 10^{-5}$ \\
$e^{\Delta S_{gg}}$ & $ 10^8$ \\
\end{tabular}
\end{center}

Collecting now the different multiplicative contributions listed 
above to the electroweak vacuum lifetime $\tau$, we  finally have 
\begin{equation}
\tau  \sim  10^{-212} \,T_U\,. \label{lifequant2}
\end{equation}

As before, we have an enormous 
difference between the tree level result (\ref{lifen}) for $\tau$ 
and the quantum corrected one (\ref{lifequant2}), but
we again see that the bulk of the contribution
to $\tau$ comes from the classical level, which, in this sense, 
provides a ``good'' estimate of $\tau$. 

In the case that we have just considered, the electroweak vacuum lifetime 
$\tau $ turns out to be enormously shorter than the age of the  
universe, thus showing that the metastability scenario is far from 
being a generic feature of theories which allow for the SM 
to be valid all the way up to the Planck scale. The expectations 
and arguments of\,\cite{isido, isiuno,isidue} are simply not fulfilled.  

Clearly, in the light of the above results, the SM phase
diagram  in the $M_H-M_t$ plane of Fig.\,\ref{bounn2}
no longer holds. For the case that we have considered, for 
instance, the instability line is tremendously lowered 
and  the big dot in the figure, corresponding to $M_H = 125.7$ 
GeV and $M_t = 173.34$ Gev, lies within the instability region. 
See \cite{bramespla}, where new phase diagrams of this kind are
plotted.    

Before ending this section, we would like to come back to the 
question of the existence of other bounce solutions and/or 
of configurations that are quasi-solutions.
In principle, if, in addition to the 
solution found above, other solutions or quasi-solutions are 
present, they could contribute to $\tau$ and the result 
(\ref{lifequant2}) should be revisited. However, this is not 
the case here. As we have just seen, in fact, the action related 
to the solution $\phi_{bou}(r)$ found above, is  
$S[\phi_{_{bou}}]\sim 80$ (see (\ref{actbounnum})), while 
for the (quasi-)solutions mentioned at the beginning of this 
section, the action is (see (\ref{actscaleinvnum}))    
$S[\phi_b]\sim 1800$. This means that the contribution of the 
latter is enormously (exponentially) suppressed as compared to 
the contribution of $\phi_{_{bou}}(r)$. 

\section{Analytical approximations}
\label{analytic}

We would like to compare now the results of the 
previous sections with those obtained in\,\cite{brames}, where the
presence of new physics interactions  was studied 
with the help of an approximation for the potential 
$V_{new}(\phi)$ in\,(\ref{potnew}) that made it possible to 
get analytic solutions for the bounces. 

\begin{figure*}[t]
\includegraphics[width=0.75\textwidth,angle=0]{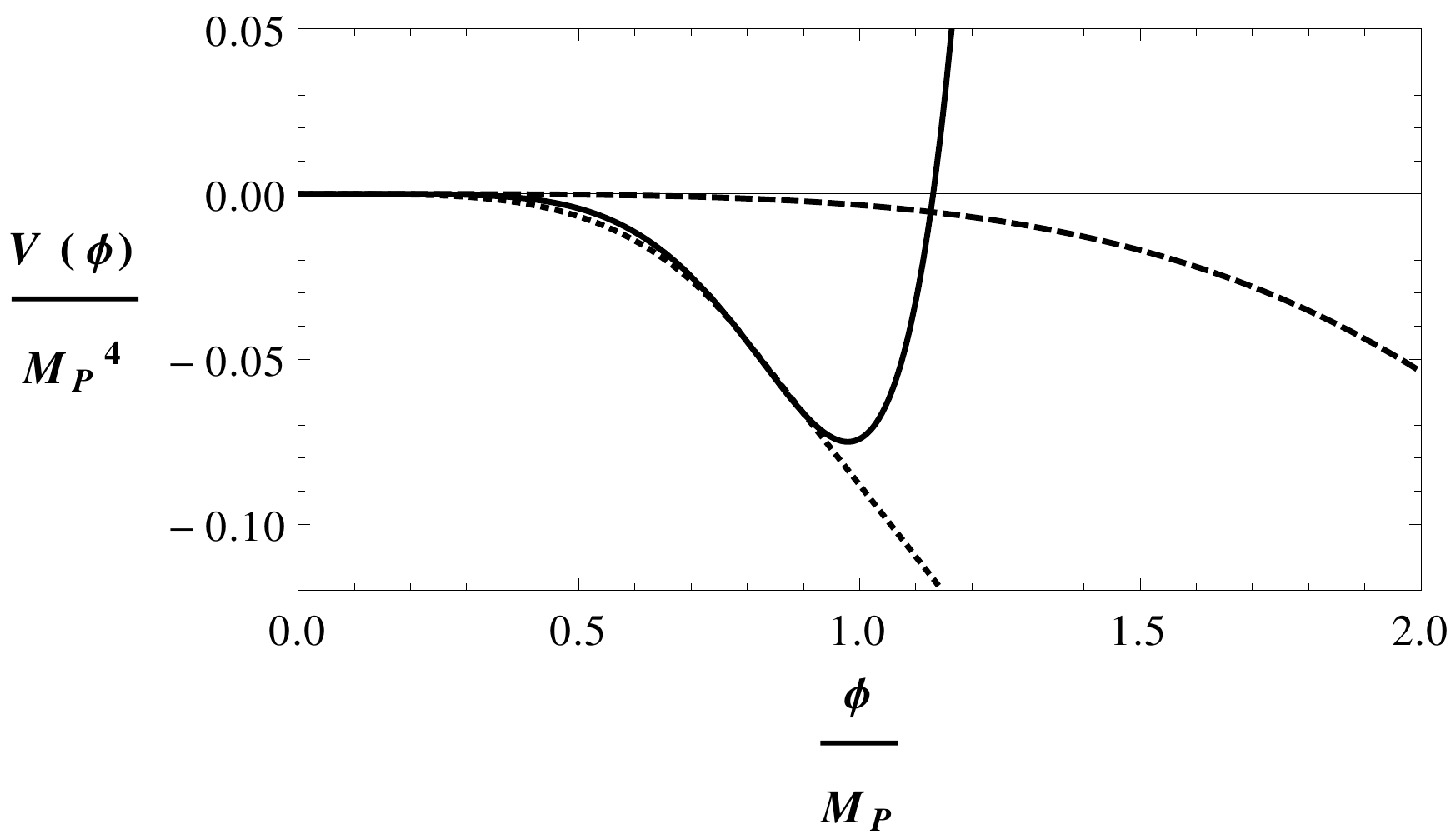}
\caption{The solid line shows the potential $V_{new}(\phi)$ 
of Eq.\,(\ref{potnew}) with $\lambda=-0.01345$, $\lambda_6=-2$ 
and $\lambda_8=2.1$. 
The dotted line is the plot of the approximation to $V_{new}(\phi)$ 
given in Eq.\,(\ref{Vapprox}), with $\eta\simeq 0.7912\, M_P$ 
(determined self-consistently in the text), 
$\lambda_{eff}=\lambda+\frac{2}{3}\lambda_6\frac{\eta^2}{M_P^2}
+\frac{1}{2}\lambda_8\frac{\eta^4}{M_P^4}=-0.4366$ and 
$\gamma=-\, \lambda_{eff} \,\,\eta^3 \, \left(\lambda\eta^3
+\lambda_6\frac{\eta^5}{M_P^2}+\lambda_8\frac{\eta^7}
{M_P^4}\right)^{-1}=-0.987$. 
As explained in the text, the latter provides a good approximation 
to $V_{new}(\phi)$ for values of $\phi$ around $\eta$. The dashed 
line is the potential in the absence of new physics interactions 
($\lambda_6=0$ and $\lambda_8=0$).} 
\label{approxpot}
\end{figure*}

The solid line in 
fig.\,\ref{approxpot} shows the plot of the potential\,(\ref{potnew})
with $\lambda=-0.01435$, $\lambda_6=-2$ and $\lambda_8=2.1$. 
Up to the scale $\eta\simeq 0.7912\, M_P$ (that will be determined 
self-consistently in the following), $V_{new}(\phi)$ 
is well approximated by an upside down quartic parabola, 
$V_{new}(\phi)\simeq\frac{\lambda_{eff}}{4}\phi^4$, with  
$\lambda_{eff}=\lambda+\frac{2}{3}\lambda_6\frac{\eta^2}{M_P^2}
+\frac{1}{2}\lambda_8\frac{\eta^4}{M_P^4}$. 
For $\phi > \eta$, $V_{new}(\phi)$ bends down creating a new 
minimum at $\phi_{min}\simeq 0.979\, M_P$. Therefore, for 
values of $\phi$ larger than (but close to) $\eta$,
$\phi \gtrsim \eta$, $V_{new}(\phi)$ can be linearized and we get
$V_{new}(\phi)=\left[\frac{\lambda_{eff}}{4}
\eta^4-\frac{\lambda_{eff} \eta^3}{\gamma}
\left(|\phi|-\eta\right)\right]$, with 
$\gamma= -\, \lambda_{eff} \,\,\eta^3 \, \left(\lambda\eta^3
+\lambda_6\frac{\eta^5}{M_P^2}+\lambda_8\frac{\eta^7}
{M_P^4}\right)^{-1}$.

The previous approximations can be included in a single expression. 
Indeed, the potential $V_{new}(\phi)$, for values 
of $\phi$ around $\eta$, can finally be written as
\begin{eqnarray}
V_{new}(\phi) \simeq \frac{\lambda_{eff}}{4}\phi^4\theta(\eta-|\phi|) 
+\left[\frac{\lambda_{eff}}{4}
\eta^4-\frac{\lambda_{eff} \eta^3}{\gamma}
\left(|\phi|-\eta\right)\right]\theta(|\phi|-\eta). \label{Vapprox}
\end{eqnarray}

The equation of motion possesses the bounce solution
\begin{eqnarray}\label{boun}
\phi_b(r)=\left\{\begin{array}{cc}
2 \eta -\eta^2 \sqrt{\frac{|\lambda_{eff}|}{8}}
\frac{r^2+\overline R^2}{\overline R} & \qquad
0<r<\overline r\\
\sqrt{\frac{8}{|\lambda_{eff}|}}
\frac{\overline R}{r^2+\overline R^2}
& \qquad
r>\overline r
\end{array}\right.
\end{eqnarray}
where 
\begin{equation}\label{erre}
\overline r=\sqrt{\frac{8\gamma}{\lambda_{eff} \eta^2}(1+\gamma)}
\,\,\,\,\,\,\,\, , \,\,\,\,\,\,\,\, 
\overline R=\sqrt{\frac{8}{|\lambda_{eff}|}\frac{\gamma^2}{\eta^2}}
\,,
\end{equation}
$\overline R$ being the size of the bounce (see Eq.(\ref{boun})),
and the action is
\begin{eqnarray}\label{bounceone}
S[\phi_b]= (1 - (\gamma + 1)^4)\,\,
\frac{8\pi^2}{3|\lambda_{eff}|}\,.
\end{eqnarray}

\begin{figure}
${}$\vskip1cm \epsfig{file=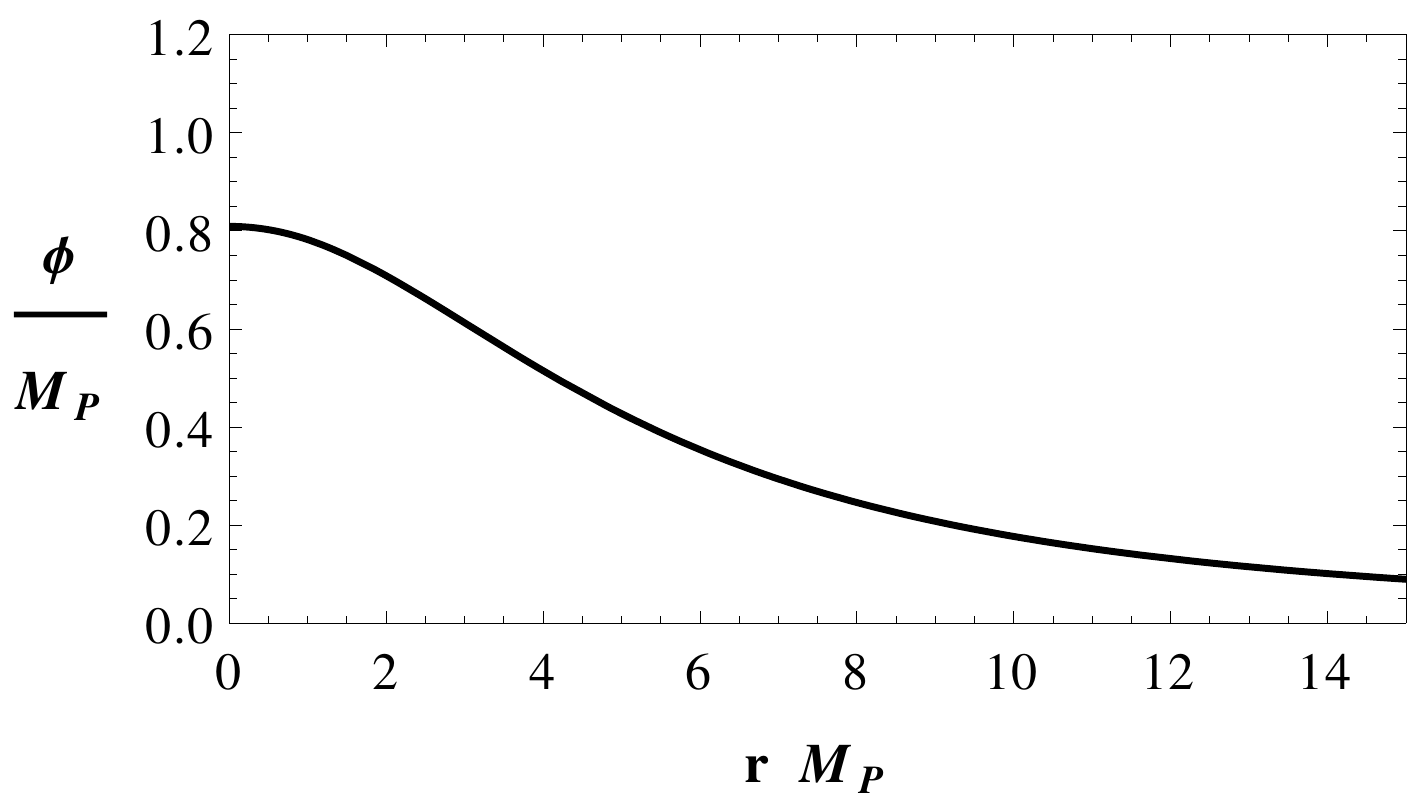,width=.75\hsize}
\caption{The analytical bounce solution, Eq.\,(\ref{boun}),
when $V_{new}(\phi)$ is approximated as in Eq.\,(\ref{Vapprox}),
for the values of the parameters considered in the text 
(see also Fig.\,\ref{approxpot}). In particular, from 
Eq.\,(\ref{erre}), we have   
$\overline r ={0.61}\,{M_P^{-1}}$, and for the bounce size,   
$\overline R={5.33}\,{M_P^{-1}}$.}\label{ban}
\end{figure}

From Eq.\,(\ref{lifen}) we see the expression for 
the main contribution to the tunneling time.
Therefore, in the approximation that we are considering, 
the tunneling time is obtained maximizing the expression
\begin{eqnarray}
{\cal T} (\eta) = \frac{\overline R(\eta)^4}{T_U^4} e^{S[\phi_b(\eta)]}
\end{eqnarray}
with respect to $\eta$.
This in turn determines the value of $\eta$ appearing in 
Eq.\,(\ref{Vapprox}).

By considering the values $\lambda=-0.01345$, 
$\lambda_6=-2$ and $\lambda_8=2.1$ of the example 
in Fig.\,\ref{approxpot}, we find $\eta =0.7912\,M_P$. 
The dotted line in this figure is the plot of the approximation
in Eq.\,(\ref{Vapprox}) for the potential $V_{new}(\phi)$
for the above value of $\eta$. We immediately see that this is 
an excellent approximation for the potential for value of 
$\phi$ close to $\eta$. In this respect, we should note that 
for the purposes of computing the bounce, this is the only region
of interest \cite{wein}. 

The profile of the bounce solution found with this approximation 
is shown in Fig.\ref{ban} and has to be compared with the bounce 
obtained numerically, shown in Fig.\ref{bouncenoi}. Moreover, 
the tunneling time under this approximation turns out to be
\be
\tau \sim 10^{-215}\,T_U,
\ee
that is a quite good estimate for $\tau$, to be compared with the 
exact numerical result of Eq.\,(\ref{lifen}). 

\section {Conclusions}

During the early discussions of the stability of the Standard Model Higgs potential, the top quark and Higgs masses were completely unknown.   It is remarkable that the values of these masses turn out to lead to a corner of parameter-space in which the stability, metastability and instability regions are so close together.   As a result, calculations need to be carried out to higher precision in order to determine the ultimate fate of our vacuum.

Although these calculations have been done, it was shown in Refs. \cite{brames,bra,bramespla} that higher dimensional Planck scale operators, neglected in previous calculations, could have an enormous effect on the tunneling rate, and thus on the lifetime of the Standard Model vacuum.    As a result, predictions of the fate of our vacuum without knowledge of these operators cannot reliably be made.

Neglecting Planck scale operators would seem to be completely reasonable, since the electroweak vacuum becomes unstable at a scale of $10^{11}$ GeV, far, far below the Planck scale.    In this paper we have pointed out two reasons why they are still important (and can dominate the tunneling rate).   First, when the Higgs field tunnels through a potential barrier (in more than one dimension), the value of the field at the center of the bubble is much, much bigger than the location of the instability.   This is because additional vacuum energy is needed to overcome the gradient terms in the Higgs Lagrangian; this is nothing other than needed a large volume energy difference to overcome surface tension.   In the SM, this results in the value of the field at the center of the bubble being roughly $10^7$ times the value at the instability, which happens to be close to the Planck scale.   Second, tunneling is an inherently non-perturbative process, and thus one's naive expectation that higher dimensional operators will have effects which are strongly Planck-scale suppressed may not be valid.   All one can do is to redo the calculations including higher dimensional operators to see if their effect is significant.   This was done in Refs. \cite{brames,bra,bramespla}, where it was shown that they can have a huge effect.

These previous calculations made several simplifying assumptions.   They used an analytic approximation to the Higgs potential and for the tunneling rate.    While this is a reasonable way to estimate the size of the Planck scale operators, a more precise calculation is needed.  In this paper, we have improved on the previous results in several ways.   We have used fully numerical techniques to solve for the bounce action and the tunneling rate, without the earlier analytic approximations.    We have included not only Higgs loop contributions to the tunneling rate, but the contributions of the other fields as well.   In addition, a toy SU(5) model shows that the type of higher dimensional operators with the given coefficients is completely reasonable.    The results confirm the earlier calculations and show that Planck scale operators do, in fact, have a huge effect on the tunneling rate.   Only with knowledge of these higher dimensional operators can the fate of our vacuum be known.

There are many other situations in which these operators can have a large effect.  As noted in Ref. \cite{bramespla}, the Higgs inflation scenario would be drastically altered.   In fact, one generally can be concerned about the basic slow-roll inflation scenario.  It is always assumed that the inflaton rolls down the potential, following the classical equations of motion.   However, while it is rolling, it could tunnel through, changing the inflation scenario completely; higher dimensional operators can drastically alter the tunneling rate, making this possibility much more likely.    Clearly, there are many potential applications of this scenario.

Finally, as the higher
dimensional Planck scale operators could have an enormous impact on 
the stability phase diagram of the Standard Model, the common 
expectation that more precise measurements of the top and Higgs masses  
would allow one to discriminate between whether our vacuum is stable or metastable (or critical)
turns out to be  unjustified. Without the knowledge of the 
(Planck scale) new physics interactions, no conclusion on the 
electroweak vacuum stability can be drawn, a better knowledge 
of $M_t$ and $M_H$ being of no help in that respect\,\cite{bramespla}. 

\vskip 20pt
{\bf Acknowledgments}
\vskip 10pt
One of us (VB) would like to thank M. Consoli, G. Degrassi, 
G. Giudice, G. Isidori, M. Krawczyk, T. Gehrmann and  
D. Zappal\`a for several helpful discussions. The work of MS was 
initiated at the Scalars 2013 Workshop in Warsaw, Poland, 
and was then supported by the National Science Foundation 
under Grant NSF-PHY-1068008.  The opinions and conclusions 
expressed herein are those of the authors, and do not 
represent the National Science Foundation.

\appendix

\vskip 20pt
\section{} \label{App2}
\vskip 15pt
In this appendix we outline the computation of the quantum 
fluctuation contribution to the electroweak vacuum lifetime 
from the different sectors of the Standard Model, see 
Equations (\ref{life}), (\ref{deltaes}) and 
(\ref{deltaS}) in the text. 

If we denote with $\chi_r(x)$ all of the SM fields 
(the index ``$r$'' indicates the different fields),
the semiclassical approximation to the path integral for
the computation of the tunneling rate is obtained 
by expanding around the configuration $\chi_r^b(x)$ 
that consists of a collection of zeroes, except for the 
case when the index $r$ indicates the Higgs field, 
in which case $\chi_r^b(x)=\phi_b(x)$, the bounce 
solution. Let us then indicate the saddle point as 
$\chi^b(x)$  

The tunneling rate is computed by performing a saddle point expansion of
the transition amplitude around $\chi^b(x)$ according to 
\begin{eqnarray}\label{ortexp}
\chi(x)=\chi^b(x)+\sum_j c_j \eta_j(x),
\end{eqnarray}
where $\eta_j(x)$ is a complete set of orthonormal 
eigenfunctions of the second variation operator 
\begin{equation}\label{seconder}
(S''[\chi_b])_{rs}=\left.\frac{\delta^2 S[\chi]}
{\delta \chi_r(x)\delta \chi_s(y)}\right|_{\chi=\chi^b},
\end{equation} 
where the $r$ and $s$ indices run over all the sectors of the 
model.

The computation of the tunneling rate is complicated by the 
presence of some zero eigenvalues in the  spectrum of the 
operator $S''[\chi^b]$ and of a negative eigenvalue.
The zero modes are related to symmetries of the classical 
action with respect to four translations (in Euclidean space-time), 
to dilatation (a symmetry that is broken by quantum effects), 
and to three $SU(2)$ global rotations. With reference to the 
two cases treated in the text, where we have considered 
the case of the Standard Model alone with the quartic 
potential, and the case where the SM is modified due to 
the presence new physics interactions, higher powers of 
the scalar field, the dilatation symmetry of the classical 
action is present only in the first case.

In the functional space, these are flat directions and we 
take care of them with the help of eight collective coordinates 
(seven in the case that the dilatation invariance is absent).  
Let us indicate with $\gamma_i$ (for $i=1,...,8$) these 
collective coordinates: the spatial coordinates $x_0^{\mu}$ 
of the center of the
bounce, the three Euler angles $\theta_i$ of the group space of 
$SU(2)$, and the size of the bounce $R$. 

Actually, the instanton (bounce) size $R$ is a collective 
coordinate only when the theory is scale invariant (dilatation 
symmetry). This is the case for the SM (when the scalar mass term 
is neglected). When new physics interactions as those appearing in
the potential (\ref{potnew}) are taken into account, the
dilatation symmetry is lost, the collective coordinate $R$
is missing, and we have only seven zero modes. 

In the following we will treat the case when all of the eight 
symmetries are present, bearing in mind that we are also 
interested to the case when dilatation symmetry is lost.
Therefore the Superdeterminant  of
the fluctuation operator is modified according to 
\begin{eqnarray}\label{modifdet}
&&\left(SDet(S''(\chi_b))\right)^{-1/2}\rightarrow \nonumber\\
&&\frac{1}{2}(2\pi)^{-8/2}\int \prod_{r=1}^8 d\gamma_r
\,\mbox{det}\left(\frac{\partial c_i}{\partial \gamma_j}\right) 
\left|SDet'(S''(\chi_b))\right|^{-1/2}.
\end{eqnarray}
where the $\gamma_i$ are the collective coordinates mentioned 
above, that allow to perform the integration
along the flat directions exactly. The contribution of 
the zero modes is encoded in the Jacobian
The factor $(2\pi)^{-8/2}$ arises to compensate the 
missing gaussian integrations 
and the negative mode provides the factor $1/2$ and the absolute 
value in the determinant \cite{cole2}.

Let us define the $SU(2)$ multiplet $\Phi_b(x)$ as

\begin{equation}
\Phi_b(x)=\frac{1}{\sqrt{2}}\left(\begin{array}{c}
0\\
\phi_b(x)
\end{array}\right),
\end{equation}

The Jacobian 
$J= (2\pi)^{-4}\mbox{det}\left(\frac{\partial c_i}
{\partial \gamma_j}\right)$ is written in terms of the 
norm of the eight linearly independent zero modes 
$\frac{\partial\Phi_b (x,\gamma)}{\partial \gamma_j}$ 
and  turns out to be
\begin{eqnarray}
J=\mbox{det}\left(\begin{array}{ccc}
\frac{1}{2\pi}\int d^4 x \, 
\partial_{\mu}\Phi_b^\dagger\partial_\nu \Phi_b & 0 & 0\\
0 & \frac{1}{2\pi} \int d^4 x\, \frac{\partial}{\partial R}
\Phi_b^\dagger 
\frac{\partial}{\partial R} \Phi_b & 0\\
0 & 0 & \frac{1}{2\pi}\int d^4 x\, \frac{\partial}
{\partial \theta_i}\Phi_b^\dagger 
\frac{\partial}{\partial \theta_j} \Phi_b
\end{array}
\right)^{1/2}.
\end{eqnarray}

Since the above matrix has a block diagonal form, 
$J$ can be expressed as the product of $J_{trans}$, 
the contribution of the translational zero modes,  
times  $J_{SU(2)}$, the contribution of the zero modes 
related to the $SU(2)$ global symmetry, times $J_{dil}$, 
the contribution of the dilatation zero mode.

The Jacobian $J_{trans}$ is given by
\begin{equation}\label{jtra}
J_{trans}= (2\pi)^{-2} \int d^4x \,\prod_{\mu=1}^4 \left[\left(\partial_\mu 
\Phi^{\dagger}_b \partial_\mu \Phi_b  
\right)\right]^{1/2}= \frac{S[\phi_b]^2}{4\pi^2}.
\end{equation}

As for the Jacobian $J_{SU(2)}$, let us consider it 
in conjunction with the integration in the three corresponding 
collective coordinates 
 
\begin{eqnarray}\label{exJu}
&&\int \prod_{r=1}^3\,d\theta_r\,\, J_{SU(2)}
=\int \prod_{r=1}^3\,d\theta_r \, 
\mbox{det}\left( \frac{1}{2\pi}\int d^4 x\, 
\frac{\partial}{\partial \theta_i}\Phi_b^\dagger 
\frac{\partial}{\partial \theta_j} \Phi_b\right)^{1/2},
\end{eqnarray}
where $\theta_1 \in [0,2\pi]$ ,  $\theta_2 \in [0,\pi]$ 
and  $\theta_3 \in [0,2\pi]$.

We can obtain an expression that is the product of a measure term 
invariant under the global $SU(2)$ transformation times a 
quantity that does not depend on the 
variables $\theta_i$. To this end, we multiply and divide 
the expression in Eq.\,(\ref{exJu}) for $\sin \theta_2$. 
Then, by further multiplying and dividing the same expression 
for $R^3$, we can also extract the dimensions from $J_{SU(2)}$
thus obtaining 
\begin{eqnarray}
\int d^3\theta\,\, J_{SU(2)}=\int d^3\theta\,
\sin\theta_2 \,\,R^3 \, J'_{SU(2)}
\end{eqnarray}
where the new dimensionless jacobian $J'_{SU(2)}$ is 
\begin{eqnarray}\label{eqnewj}
J'_{SU(2)}=\frac{1}{R^3 \sin \theta_2} 
\mbox{det}\left( \frac{1}{2\pi}\int d^4 x\, 
\frac{\partial}{\partial \theta_i}\Phi_b^\dagger 
\frac{\partial}{\partial \theta_j} \Phi_b\right)^{1/2},
\end{eqnarray}
and the invariant measure is $d^3\theta\,\sin\theta_2$.
 
$J'_{SU(2)}$ can be now be made explicit by writing  
$\Phi_b$  in terms of a generic $SU(2)$ transformation applied 
to $\Phi_b^0$ defined as $\Phi_b^0\equiv
\Phi_b(x,x_0,R,\theta_i=0)$. 
By replacing then
\begin{equation}
\Phi_b=e^{i\theta_1 T_1}e^{i\theta_2 T_2}e^{i\theta_3 T_3}\Phi_b^0 
\end{equation} 
in Eq.\,(\ref{eqnewj}) and performing some algebraic manipulations we get

\begin{eqnarray}\label{jsu2}
 J'_{SU(2)}=\frac{1}{R^3}\det\left(\frac{1}{2\pi}\int d^4x \Phi_b^{0\dagger} T_i^\dagger \cdot T_j \Phi_b^0 \right)^{1/2}
= \frac{1}{R^3}\left[\frac{1}{2\pi}\int d^4x\, \phi_b^2\right]^{3/2},
\end{eqnarray}  
where $T_i$ (for $i=1,2,3$) is the real
 representation of the $SU(2)$ generators.

Finally  the contribution of the  dilatational zero mode  $J_{dil}$ is 
\begin{equation}\label{jdil}
J_{dil}= \left(\frac{1}{2\pi}\int d^4x\, \left(\frac{\partial \phi_b}{\partial R}\right)^{2}\right)^{1/2}.
\end{equation}

Bearing in mind that the integration over the $SU(2)$ angular 
variables provides a factor $16\pi^2$ and that the volume factor 
$\int d^4x_0$ is four times the time of the universe $T_U$, 
referring to  Eq.\,(\ref{tunnel}) in the text, we find that
the tunneling rate $T_U/\tau$ for unit volume and time is
\begin{eqnarray}\label{tunnelingform}
p=e^{-S[\chi_b]} 16\pi^2 \, V\,T_U \int dR 
\,R^3\, J_{trans} J_{SU(2)} J_{dil} \left|
\frac{SDet'(S''(\chi_b))}{SDet(S''(0))}\right|^{-1/2}.
\end{eqnarray}
where $T_U$ is the age of the universe and $V$ the volume 
($V=T_U^3$). 
Note that the  dimensional factor $T_U^4 \int dR \,R^3\,$ is 
compensated by the 
dimension of the ratio ${SDet'(S''(\chi_b))}/{SDet(S''(0))}$.

Finally, we recall that the fluctuation determinant breaks the 
scale invariance, so that only one of the bounces, with a 
specific value of the size $R$, dominates the above integral. 
Referring again to the notation introduced in the text, we 
indicate with $R_M$ this value of $R$ and we have
\begin{eqnarray}\label{tunnelingtau}
\frac{\tau}{T_U}= \frac{R^4_M}{T_U^4} e^{S[\phi_b]} 
\left(\frac{16\pi^2}{R^8} J_{trans} J_{SU(2)} J_{dil} 
\left|\frac{SDet'(S''(\phi_b))}{SDet(S''(0))}\right|^{-1/2}
\right)^{-1}_{R=R_M}\,,
\end{eqnarray}
that immediately brings to Eq.\,(\ref{deltaes}) used in the text.

It is worth stressing here that when the dilatation
symmetry is absent, as is the case for the modified potential
considered in this paper, where new physics interactions are added 
to the usual SM potential (see Eq. (\ref{potnew}) in the text),
the above formula has to be modified in the following three aspects.
The size $R$ of the bounce that appears in (\ref{tunnelingtau}) 
is no longer the result of the maximization of the integrand
function, but comes directly from the equation of motion (the 
action is not scale invariant already at the classical level, 
so we have only one bounce, no degeneracy). For the same reason, 
$J_{dil}$ is absent and the factor $R^{-8}$ becomes $R^{-6}$.

The next step concerns the evaluation of the ratio
\begin{eqnarray}\label{fluctua}
\left|\frac{SDet'(S''(\chi_b))}{SDet(S''(0))}\right|^{-1/2},
\end{eqnarray}
with contributions from the different 
sectors of the Standard Model. More specifically, we have to 
compute the contribution from the Higgs field $\phi$, 
the three Goldstone bosons $G_i$ (for $i=1,2,3$), 
the four gauge fields $A_{\mu}^a$ (for $a=1,2,3,4$), 
the four corresponding ghost fields $c_a$ and 
the heaviest matter contribution, i.e. the contribution 
from top quark $\psi$ (the contribution of the other fermion 
fields are far  less important and can be neglected).

In the following we will see that the $S''$ operator takes 
 block diagonal form, each block being related to one of the 
following three different sectors: Higgs, top, and 
gauge + goldstone.
To this end, we write down the different contribution to 
the $EW$ Lagrangian and extract its quadratic part in the 
fields, the only part that is relevant for the computation 
of the fluctuations around the bounce.

The action of the scalar sector of the model, 
Eq.\,(\ref{actionSec}) is usually written in terms of 
the $SU(2)$ doublet of Eq.\,(\ref{doublet}) 
(here we write $\phi=\phi_b + H$)
\begin{equation}
\Phi=\left(\begin{array}{c} \phi^+ \\ \phi^0
\end{array}\right)=\frac{1}{\sqrt{2}}\left(\begin{array}{c}
-i(G_1-iG_2)\\
\phi_b + H+i G_3
\end{array}\right)\,.
\end{equation}

However for our purposes it is useful to consider the real 
four dimensional representation of the $SU(2)\times U(1)$ group 
acting on the scalar multiplet $\phi_i =(G_1,G_2,G_3, \phi_b + H)$,
so that by adding the interaction term between the scalars and 
the gauge fields we get
\begin{eqnarray}\label{scalarLagr}
\mathcal{L}_{scalar}&=&\frac{1}{2}
\left(D_\mu \phi_i\right)^2+ V(\phi_i ^2 )\nonumber\\
&=&\frac{1}{2}\left(\partial_\mu \phi_i\right)^2+ V(\phi_i ^2 )
+\frac{1}{2} g^2_a(T^a)_{ji}(T^b)_{jk}  
\phi_i\phi_k A^a_\mu A_\mu^b\nonumber\\
&+&g_a (T^a)_{ij}\partial_\mu \phi_i \,\phi_j A^a_\mu 
\end{eqnarray}
where (with the mass term neglected, i.e. for large values 
of the scalar field)
\begin{equation}\label{potentialvecchio}
V(\phi_i^2)=\frac{\lambda}{4}(\phi_i\phi_i)^2
\end{equation}
when we consider the SM interactions only. When, on the contrary, 
we also take into account the presence of new physics 
interactions as those considered in Eq.\,(\ref{potnew}),
the potential takes the form 
\begin{equation}\label{potentialnuovo}
V(\phi_i^2)=\frac{\lambda}{4}(\phi_i\phi_i)^2 +
\frac{\lambda_6}{6\,M_P^2}(\phi_i\phi_i)^3 +
\frac{\lambda_8}{8\,M_P^4}(\phi_i\phi_i)^4\,.
\end{equation}

The computation of the fluctuation determinant in the 
presence of these additional terms presents quite nontrivial 
aspects. However, for the time being, we continue to write 
the formulas referring only to the potential of 
Eq.\,(\ref{potentialvecchio}), bearing in mind 
that they have to be modified by inserting the potential 
Eq.\,(\ref{potentialnuovo}) when we take into 
account the presence of new physics.

Note that in Eq.\,(\ref{scalarLagr}) we have written the 
covariant derivative $D_\mu$  in terms of the $4\times 4$ 
$SU(2)\times U(1)$ generators $T^a$ ($a=1,2,3,4$), 
of the four gauge bosons  $A_\mu^a$ and of the  gauge coupling 
$g_a$ (that are $g$ for $a=1,2,3$ and $g'$ for $a=4$, i.e. 
the usual $SU(2)$ and $U(1)$ couplings, respectively) as
\begin{equation}
D_\mu=\partial_\mu +g_{a}  T^a A^a_\mu.
\end{equation}

The quadratic part of Eq.\,(\ref{scalarLagr}) is therefore given by
\begin{eqnarray}\label{scal2part}
\mathcal{L}^{(2)}_{scalar}&=&\frac{1}{2}(\partial_\mu \phi_b)^2+\frac{\lambda}{4}\phi_b^4+
\frac{1}{2}(\partial_\mu H)^2+ \frac{3}{2}\lambda \phi_b^2 H^2\nonumber\\
&+&\frac{1}{2}\sum_i(\partial_\mu G_i)^2+\frac{\lambda}{2}\phi_b^2\sum_i G_i^2\nonumber\\
&+&\frac{\phi_b^2}{8}\left(g^2 A^{1\mu}A^{1}_\mu + g^2 A^{2\mu}A^{2}_\mu
+(g^2+g^{'2}) Z^{\mu}Z_\mu\right)\nonumber\\
&+& g A^1_\mu \phi_b\partial_\mu G_1+g  A^2_\mu \phi_b\partial_\mu G_2+ \sqrt{g^2+g^{'2}}
Z^\mu \phi_b\partial_\mu G_3\nonumber\\
&+&\frac{g}{2}\partial^\mu A^1_\mu \phi_b G_1 +\frac{g}{2}\partial_\mu A^{2\mu} \phi_b G_2
+\frac{\sqrt{g^2+g^{'2}}}{2}\partial_\mu Z^\mu \phi_b G_3
\end{eqnarray}
where the equation of motion $-\partial^2 \phi_b +\lambda \phi_b^3=0$ 
has been used and 
we have rotated the gauge field $A^3_\mu$ and $A^4_\mu$ according to 
the transformations
\begin{eqnarray}\label{Arotations}
A^3_\mu &=&\frac{1}{\sqrt{g^2+g'^2}}(g Z_\mu + g' A_\mu)\nonumber\\
A^4_\mu &=&\frac{1}{\sqrt{g^2+g'^2}}(g A_\mu - g' Z_\mu)\,.
\end{eqnarray}

The kinetic term for the four gauge bosons is given by
\begin{equation}\label{gaugelag}
\mathcal{L}_{gauge,kin}= \frac{1}{4}F^a_{\mu\nu}F^{a\mu\nu}
\end{equation}
where 
\begin{equation}
F^a_{\mu\nu}=\partial_\mu A_\nu^a - \partial_\nu A_\mu^a+ g_a f^{abc}A^b_\mu A^c_\nu.
\end{equation}

The $f^{abc}$ are the  structure constants of the group which are equal to 
$\epsilon^{abc}$  when all the indices take one of the values $1,2,3$ 
and zero otherwise.
The quadratic part in the gauge fields of the lagrangian in Eq.\,(\ref{gaugelag}) 
is given by
\begin{eqnarray}
\mathcal{L}^{(2)}_{gauge,kin}&=&\frac{1}{2}\sum_{a=1}^4 A^a_\mu \left(-\partial^2 \delta^{\mu\nu}+\partial^\mu \partial^\nu\right) A^a_\nu
=\frac{1}{2}\sum_{i=1}^2 A^i_\mu \left(-\partial^2 \delta^{\mu\nu}+\partial^\mu \partial^\nu\right) A^i_\nu \nonumber\\
&+&\frac{1}{2} A_\mu \left(-\partial^2 \delta^{\mu\nu}
+\partial^\mu \partial^\nu\right) A_\nu 
+\frac{1}{2} Z_\mu\left(-\partial^2 \delta^{\mu\nu}
+\partial^\mu \partial^\nu\right)Z_\nu,
\end{eqnarray}
where again the rotation in Eq.\,(\ref{Arotations}) is considered.

We use the $R_{\xi}$ gauge fixing, so that  
the gauge fixing lagrangian is written as
\begin{equation}\label{fixlag}
\mathcal{L}_{gauge,fix}=\frac{1}{2\xi}\left(\partial_\mu A^{a\mu}+
\xi g_a (T^a)_{ij}\phi_b^j \left(\phi^i-\phi_b^i\right)
\right)^2.
\end{equation}

The quadratic part of the lagrangian in Eq.\,(\ref{fixlag}) 
is 
\begin{eqnarray}\label{2laggauge}
\mathcal{L}^{(2)}_{gauge,fix}&=&-\frac{1}{2\xi}\sum_{i=1}^2 A^i_\mu \partial^\mu \partial^\nu A^i_\nu
-\frac{1}{2\xi} A_\mu \partial^\mu \partial^\nu A_\nu-\frac{1}{2\xi} Z_\mu \partial^\mu \partial^\nu Z_\nu\nonumber\\
&+&\frac{\xi}{8}\phi_b^2\left(g^2(G_1^2+G_2^2)+(g^2+g^{'2})G_3^2\right)\nonumber\\
&+&\frac{g}{2}\partial^\mu A^1_\mu \phi_b G_1 +\frac{g}{2}\partial_\mu A^{2\mu} \phi_b G_2
+\frac{\sqrt{g^2+g^{'2}}}{2}\partial_\mu Z^\mu \phi_b G_3.
\end{eqnarray}

Note that the terms that mix the gauge and Goldstone 
fields in Eq.\,(\ref{2laggauge}), together with the analogous 
terms in  Eq.\,(\ref{scal2part}), give
\begin{eqnarray}
-g A^1_\mu \partial^\mu \phi_b G_1 -g\partial_\mu A^{2\mu} \phi_b G_2
-\sqrt{g^2+g^{'2}} Z^\mu \partial_\mu \phi_b G_3.
\end{eqnarray}

Moreover, the contribution to the determinant coming from 
the field $A_\mu$ in Eqs.\,(\ref{scal2part}) and (\ref{2laggauge})
is the same as in the free case. Therefore, 
when the ratio of determinants is performed, this terms disappear.

In addition to the gauge fixing terms, the Fadeev-Popov 
quantization also requires 
the introduction of four additional ghost fields $c^a$ 
(with the corresponding conjugate fields $c_a^*$), the 
lagrangian being
\begin{equation}\label{ghost}
\mathcal{L}_{ghost}=c_a^*
\left[-\partial^\mu D^{ab}_\mu+\xi g^2_a (T^a\cdot\phi_b)\cdot
(T^b\cdot\phi)\right]c_b,
\end{equation} 
where the covariant derivative for the ghost fields is given by
$ D_\mu^{ac}=\partial_\mu \delta^{ac}+g_a f^{abc}A^b_\mu$. 
The quadratic part of (\ref{ghost}) is 
\begin{eqnarray}
\mathcal{L}_{ghost}^{(2)}&=&\sum_{i=1}^2 c_i^* 
\left(-\partial^2+ \xi \frac{g^2}{4}\phi_b^2 \right) c_i
+ c_3^* \left(-\partial^2+ \xi \frac{g^2+g^{'2}}{4}\phi_b^2\right) c_3\nonumber\\
&+& c_4^*\left(-\partial^2\right) c_4.
\end{eqnarray}
As in the case of the  $A_\mu$ fields above, the ghost 
$c_4$ gives the same contribution
as in the free case, then it can be neglected.

Finally, for the fermions fields, the only relevant contribution
comes from the top quark (all the other contributions being 
negligible). The quadratic part of the 
top lagrangian, in the bounce background field, is then ($g_t$
is the Yukawa top coupling and $\psi$ the top field)

\begin{eqnarray}
\mathcal{L}^{(2)}_F=\bar \psi \left(\slashed \partial +\frac{g_t}
{\sqrt{2}}\,\phi_b\right)\psi.
\end{eqnarray}

With all the above building blocks at our disposal, we are 
finally in the position to write the fluctuation operator 
$S''(\chi_b)$. It takes the block diagonal form
\begin{eqnarray}
S''(\chi_b)=\left(
\begin{array}{ccccccc}
S_{H H} & 0 & 0 & 0 & 0 & 0&0 \\
0 & 0 & S_{\psi\bar\psi} & 0 & 0& 0&0\\
0 & S_{\bar\psi \psi} & 0 & 0 & 0& 0&0\\
0&0&0&S_{\tilde A^{i} \tilde A^{i}}& S_{\tilde A^i G^i}& 0&0 \\
0&0&0&S_{G^i \tilde A^i}&S_{G^i G^i}& 0&0\\
0 & 0 & 0 & 0 & 0& 0&S_{c_i c_i^*}\\
0 & 0 & 0 & 0 & 0& S_{c_i^* c_i}&0
\end{array}\right)\label{deltaesse}
\end{eqnarray}
where $i=1,2,3$ and we have set 
$\tilde A^{i}_\mu=(A^1_\mu, A^2_\mu, Z_\mu)$. 

Since this matrix is block diagonal, $SDet$ in Eq.(\ref{fluctua}) 
becomes the product of the different determinants appearing
in the different blocks, i.e. the product of the determinants of
the operators
\begin{eqnarray}
S''_H&\equiv&S_{HH}\nonumber\\
S''_t&\equiv& \left(\begin{array}{cc}
0& S_{\psi\bar\psi} \\
 S_{\bar\psi \psi} &0
\end{array}\right)\nonumber\\
S''_{gg}&\equiv&\left(\begin{array}{cc}
S_{\tilde A^{i} \tilde A^{i}}& S_{\tilde A^i G^i} \\
S_{G^i \tilde A^i}&S_{G^i G^i}\end{array}\right)\\
S''_{ghost}&\equiv& \left(\begin{array}{cc}
 0&S_{c_i c_i^*}\\
 S_{c_i^* c_i}&0
\end{array}\right)
\end{eqnarray}  

We can then write the tunneling time in 
Eq.(\ref{tunnelingtau}) as
\begin{eqnarray}\label{tunnelingt}
\frac{\tau}{T_U}= \frac{R^4_M}{T_U^4} e^{S[\phi_b]} 
e^{\Delta S_H + \Delta S_t + \Delta S_{gg}} 
\end{eqnarray}
where 
\begin{eqnarray}
\Delta S_H&=&\frac{1}{2}\ln\left(\frac{1}{R_M^{10}}\frac{\mbox{Det}'S''_H[\phi_b]}{\mbox{Det}S''_H[0]}\right)-\ln J_{trans}-\ln J_{dil}
\label{deltash}\\
\Delta S_t&=&-\frac{3}{2}\ln\left(\frac{\mbox{Det}S''_t[\phi_b]}{\mbox{Det}S''_t[0]}\right)\label{deltast}\\
\Delta S_{gg}&=&\frac{1}{2}\ln\left(\frac{1}{R_M^{6}}\frac{\mbox{Det}'S''_{gg}[\phi_b]}{\mbox{Det}S''_{gg}[0]}\right)-\frac{1}{2}\ln\left(\frac{\mbox{Det}S''_{ghost}[\phi_b]}{\mbox{Det}S''_{ghost}[0]}\right)
-\ln (16\pi^2 J_{SU(2)})\label{deltasgg}\,. 
\end{eqnarray}
Eq. (\ref{tunnelingt}) has to be compared with 
Eq.\,(\ref{life}) in the text. 

It is important to note that the contribution $\Delta S_H$ of 
Eq.\,(\ref{deltash}) is greatly modified when the potential 
with the new physics interactions (\ref{potentialnuovo})
replaces the SM potential (\ref{potentialvecchio}). Namely,
$J_{dil}$ is missing and $R^8$ rather than $R^{10}$ appears
(we have already commented on the size of the bounce to be 
considered). 

Let us compute  the different contributions to the 
fluctuation determinant, 
(\ref{deltash}), (\ref{deltast}), and (\ref{deltasgg}), 
in the two cases of interest for us, 
namely the case where only SM interactions are considered,
potential given by 
Eq.\,(\ref{potentialvecchio}) (Section II), and the case 
where we take 
into account the new physics interactions at the Planck 
scale, namely the case of the potential 
\,(\ref{potentialnuovo}) (Section III). 

Let us begin with the Jacobian factors. As for 
$J_{trans}$, that appears in Eq.\,(\ref{deltash})
for $\Delta S_H$, 
from Eq.\,(\ref{jtra}) we already know that 
\be\label{lnjt}
-\ln J_{trans}=-\ln \frac{S[\phi_b]^2}{4\pi^2}\,.
\ee
In the case of the SM potential alone (Section II), 
Eq.\,(\ref{potentialvecchio}), we have (see Eq.\,(\ref{actscaleinv}))
\be
-\ln J_{trans}^{SM}=-\ln\frac{16\pi^2}{9\lambda^2} . 
\ee
Inserting the value of $\lambda$ considered in the text
($\lambda=-0.01345$), we get
\be \label{jtransnumSM}
-\ln J_{trans}^{SM} \sim -11.5\,. 
\ee

If we now consider the potential with the inclusion of the
new physics interactions (Section III), while $\ln J_{trans}$ is
still given by Eq.\,(\ref{lnjt}), we no longer have an analytical 
expression for $S[\phi_b]$. In fact, we compute the bounce 
solution $\phi_b(x)$ numerically in the next appendix, 
so that in turn we obtain $S[\phi_b]$ numerically. For the 
values of $\lambda$, $\lambda_6$ and $\lambda_8$ considered 
in the text (see Section III), we have
\be \label{jtransnumnew}
-\ln J_{trans}^{new} \sim -5.14\,. 
\ee

Let us consider now the contribution of $J_{dil}$ to $\Delta S_H$. 
As  we have already said, the contribution of $J_{dil}$ appears 
only for the SM case. From Eq.\,(\ref{jdil}) we see that this 
contribution is given by 
\bea
-\ln J_{dil}^{SM}&=&-\frac{1}{2} \ln
\left(\frac{1}{2\pi}\int d^4x\, 
\left(\frac{\partial \phi_b}{\partial R}\right)^{2}\right)
= -\frac{1}{2}\ln\left(\frac{8\pi^2}{|\lambda|}
\int^{\frac{1}{R_M v}}_0 dy \,y^3
\frac{(y^2-1)^2}{(1+y^2)^4}\right)\nonumber\\
&=&-\frac{1}{2}\ln\left(\frac{8\pi^2}{|\lambda|}\ln\frac{1}{R_M v}
\right)\, , 
\eea
where we have defined $y$ as $y=r/R_M$. Moreover, the integral over 
the radial coordinate $r$ is infra-red divergent. This is due to 
the fact that in the potential the mass term has been neglected. 
For this reason, an infra-red cut-off $r=1/v$ has been inserted, 
thus getting the above result. By considering the values of 
$\lambda$ and $R_M$ given in the text, we get
\be \label{jdilnumSM}
-\ln J_{dil}^{SM}= -6.07\,. 
\ee

Finally we move to the contribution of $J_{SU(2)}$ to 
$\Delta S_{gg}$. From Eq.\,(\ref{jsu2}) we have
\be\label{jsu22}
-ln (16\pi^2\, J_{SU(2)})=-\frac{3}{2}\ln\left((16\pi^2)^{2/3}
\left[\frac{1}{2\pi}\int d^4x \frac{\phi_b^2(r)}{R_M^2}\right]\right)\,.
\ee
In the case of the SM potential alone (Section II), 
Eq.\,(\ref{potentialvecchio}), we have 
\bea
-ln (16\pi^2\, J_{SU(2)})&=&-\frac{3}{2}
\ln\left(\frac{2^{17/3}\pi^{7/3}}
{|\lambda|}\int^{\frac{1}{R_M v}>>1}_{0} dy
\frac{y^3}{(1+y^2)^2}\right)\nonumber\\
&= &-\frac{3}{2}\ln\left(\frac{2^{17/3}\pi^{7/3}}
{|\lambda|}\ln\frac{1}{R_M v}\right)\,
\eea
where, as for $J_{dil}$, $y=r/R_M$ and we have inserted 
an infra-red cut-off $r=1/v$. By considering the values of 
$\lambda$ and $R_M$ given in the text, we get
\be \label{jsu2numSM}
-ln (16\pi^2\, J_{SU(2)}^{SM})=-22.6\,.
\ee

If we now consider the potential with the inclusion of the
new physics interactions (Section III), as for the case of  
$J_{trans}$, we have to move to the numerical evaluation of
the bounce solution (Section III and Appendix \ref{App3}). 
Then, by taking the values of $\lambda$, $\lambda_6$ and 
$\lambda_8$ considered in the text (see Section III), from 
Eq.\,(\ref{jsu22}) we get
\be \label{jsu2numnew}
-\ln J_{SU(2)}^{new} \sim -15.4\,. 
\ee

Let us move now to the computation of the determinants, 
and focus our attention on $\Delta S_H$, i.e.  
on $S''_H$. As is well known, the functional determinant is 
obtained by solving the eigenvalue equation
\be
S''_H \psi = \lambda \psi,
\ee
where $\psi$ are the eigenfunctions of $S''_H$ and $\lambda$ the
corresponding eigenvalues. In $\Delta S_H$, the ratio 
${\mbox{Det}'S''_H[\phi_b]}/{\mbox{Det}S''_H[0]}$ appears.
The prime in the determinant is due to the fact that only 
the non zero eigenvalues have to be considered in the 
evaluation of the determinant.

As $S''_H(\phi_b)=-\partial^2+V''(\phi_b)$, we have to  
compute 
\be
\frac{\rm det' (-\partial^2+V''(\phi_b))}
{\rm det (-\partial^2)}\,.
\ee
Due to radial symmetry, $V''(\phi_b)$ in 
$\left[-\partial^2+V''(\phi_b)\right]$ only depends on $r$, 
and we can use the powerful Gelfand-Yaglom method 
for the computation of the determinant. 
Following \cite{yaglom}, the 
logarithm of the ratio of determinants, with some specifications 
given below, is then 
obtained as ($j=0, 1/2, 1, 3/2, 2,...$)  
\be \label{yaglom}
\log\left(\frac{\rm det' (-\partial^2+V''(\phi_b))}
{\rm det (-\partial^2)}\right)^{1/2}=
\frac{1}{2}\sum_{j= 0}^{\infty} (2j+1)^2\ln \rho_j\\
\ee 
\be\label{limitrho}
{\rm where}\,\,\,\,\,\,\rho_j =\lim_{r\rightarrow \infty} 
\rho_j(r)\,\,\,\,\,\,\,\,\,\,\,\,\,\,\,\,\,\,\,\,\,\,\,\,
\ee
and each of the $\rho_j(r)$ is solution of the differential equation
\begin{eqnarray}\label{yagloeq}
\rho_j''(r)+\frac{\left(4j+ 3\right)}
{r}\rho_j'(r)-V''(\phi_b(r))\rho_j(r)=0
\end{eqnarray}  
with boundary conditions $\rho_j(0)=1$ and $\rho_j'(0)=0$.
($\rho_j''(r)$ is the second derivative of $\rho_j(r)$ w.r.to $r$,...).
As for the laplacian operator $\partial^2$, we can write it as
\be\label{jei}
\partial^2=\frac{d^2}{dr} + \frac{3}{r}\frac{d}{dr}
-\frac{\hat J^2}{r^2}\, ,
\ee
where the operator $\hat J^2$ is 
$\hat J^2 = \hat J_{\mu\nu} \hat J_{\mu\nu}$, with $\hat J_{\mu\nu}=
-\frac{i}{\sqrt{2}}(x_\mu\partial_\nu-x_\nu\partial_\mu)$, 
``angular momentum operator'' in $R^4$. The eigenfunctions of 
of $J^2$ are the hyperspherical harmonics $Y_j^{m,m'}$  
($m,m'= -j,..., +j$) and the eigenvalues are $\lambda_j= 4 j(j+1)$,
with degeneracy $(2j+1)^2$. Each of the $\rho_j$ is the 
product of eigenvalues of the operator 
$S''_H(\phi_b)=-\partial^2+V''(\phi_b)$ divided by the product 
of eigenvalues of $\partial^2$,  where the operator $\hat J^2$ of 
Eq. (\ref{jei}) is replaced by the eigenvalue $4j(j+1)$.

Eq.\,(\ref{yaglom}) is ill defined in the following three aspects. 
One of the eigenvalues related to $j=0$ is negative, 
and a second one is vanishing and is related to the 
dilatation invariance of the theory. Actually, this is true 
only when we do not consider the presence of new physics 
interactions, in which case there is no dilatation invariance. Moreover, four of 
the eigenvalues entering 
in $\rho_{1/2}$ vanish, as they correspond to the four translational 
zero modes. Actually $\rho_0$ and $\rho_{1/2}$ can be separately 
treated in a standard way\,\cite{yaglom,dunne} (see below). 
Finally, the sum in 
Eq.\,(\ref{yaglom}) is divergent. This is the usual UV 
divergence. 

If we consider, for instance, the SM case with the $\lambda\phi^4$ 
potential, inserting the bounce, Eq. (\ref{bouncecla}), in $V''(\phi_b)$
of Eq.\,(\ref{yagloeq}), and then taking the limit in 
Eq. (\ref{limitrho}) 
we have
\be
\rho_j =\frac{j(2j-1)}{(j+1)(2j+3)}
\ee

From the above equation, it is immediate to see that, if we cut 
the sum in Eq.\,(\ref{yaglom}) to a maximal 
value of $j$, say $j=j_{max}$, we get
terms proportional to $j_{max}$ (quadratic divergences),
terms proportional to ${\rm ln}\, j_{max}$ (logarithmic divergences), 
finite terms and then terms $O(1/j_{max})$. 

If we now consider the potential with the insertion of the new 
physics operators, Eq.\,(\ref{potentialnuovo}), the differential
equations (\ref{yagloeq}) can be solved only numerically. However,
also in this case, we can still easily recognize 
the quadratic and logarithmic divergences as well as the finite 
contributions. 

In order to get rid of these divergences, we have to follow 
the usual 
renormalization procedure, i.e. we have to introduce 
counterterms $\delta S_H^{ct}$, and get for the renormalized sum   
\begin{eqnarray}\label{sumrenorm}
\left[\frac{1}{2}\sum_{j=0}^{\infty} (2j+1)^2\ln \rho_j\right]_r
\equiv
\frac{1}{2}\sum_{j=0}^{\infty} (2j+1)^2\ln \rho_j
-\delta S_H^{ct}\,.
\end{eqnarray}

Naturally, the determination of the counterterms depends on 
the choice of the renormalization conditions and scheme.  
One possibility consists in extracting the divergences from 
Eq. (\ref{yaglom}) by expanding the $\rho_j$
for large values of $j$. The first two terms 
of this expansion provide nothing but  the 
quadratic and logarithmic divergences. By subtracting these 
terms, we operate a specific choice of counterterms 
$\delta S_H^{ct}$, that finally would  lead to renormalized 
quantities, in particular to the renormalized quartic coupling. 

However, in order to make contact with the existing literature, 
it is convenient to adopt a more conventional renormalization 
procedure, namely the ${\overline {MS}}$ scheme. This amounts to 
the following procedure \cite{isido}. 

First we solve perturbatively the differential 
equation for the $\rho_j(r)$, Eq.\,(\ref{yagloeq}), by 
considering $V''(\phi_b)$ as a perturbation, expanding
the functions $\rho_j(r)$ as $\rho_j(r)=1+\rho_j^{(1)}(r)+
\rho_j^{(2)}(r) +\cdots $, and assuming $\rho_j^{(1)}(r) \sim 
\mathcal{O}\left(V''(\phi_b)\right)$ and $\rho_j^{(2)}(r) \sim 
\mathcal{O}\left(V''(\phi_b)^2\right)$. Then we take the limit 
for $r\to \infty$ and compute the  expression
\begin{equation}\label{theroseries}
\sum^{\infty}_{j=0}(2j+1)^2\left(\ln\rho_j-\rho_j^{(1)}
+\frac{1}{2}(\rho_j^{(1)})^2-\rho_j^{(2)}\right)
\end{equation}
which turns out to be finite.
This is because the above combination of $\rho^{(1)}$ and 
$\rho^{(2)}$ has the same divergences of $\ln\rho_j$.
Referring again to Eq.\,(\ref{yaglom}), one immediately 
verifies that such a  procedure corresponds to subtract
from the first member of Eq.\,(\ref{yaglom}) the first
two terms of the perturbative expansion
\begin{eqnarray}\label{diver2}
\frac{1}{2}\mbox{Tr}\ln\left[1+(-\partial^2)^{-1}
V''(\phi_b)\right]
&=&\frac{1}{2}\mbox{Tr}\left[(-\partial^2)^{-1}
V''(\phi_b)\right]\nonumber\\
&-&\frac{1}{4}\mbox{Tr}\left[(-\partial^2)^{-1}
V''(\phi_b)(-\partial^2)^{-1}V''(\phi_b)\right]\nonumber\\
&+&\mathcal{O}\left((V'')^3\right)\,.
\end{eqnarray} 

Finally, the contact with existing literature is made when
Eq.\,(\ref{sumrenorm}) is written by adding and subtracting
the quadratic and logarithmic divergencies written once in the 
form given in Eq.\,(\ref{theroseries}), once in the form 
given in Eq.\,(\ref{diver2}), i.e. by writing

\bea\label{sumsum}
&&\left[\frac{1}{2}\sum_{j=0}^{\infty} (2j+1)^2\ln \rho_j\right]_r
=\sum^{\infty}_{j=0}(2j+1)^2\left(\ln\rho_j-\rho_j^{(1)}
+\frac{1}{2}(\rho_j^{(1)})^2-\rho_j^{(2)}\right)\nonumber\\
&+&\frac{1}{2}\mbox{Tr}\left[(-\partial^2)^{-1}
V''(\phi_b)\right]
-\frac{1}{4}\mbox{Tr}\left[(-\partial^2)^{-1}
V''(\phi_b)(-\partial^2)^{-1}V''(\phi_b)\right]-
\delta S_H^{ct}
\eea

The sum in the r.h.s of the first line is computed numerically.
For the potential in Eq.\,(\ref{potentialvecchio}), i.e. for the 
potential of the SM alone, the result 
does not depends on the values of the $SM$ couplings. By 
performing the numerical computation for this sum, we get: $6.02$.  
When we include the couplings $\lambda_6$ and
$\lambda_8$, i.e. when we consider the potential of 
Eq.\,(\ref{potentialvecchio}), we find that the 
sum  depends on these latter couplings
as well as on the other ones. For the numerical example considered 
in the text, $\lambda_6=-2$ and $\lambda_6=2.1$, and for 
the central values of the top and Higgs masses, 
$M_t=173.34$ GeV and $M_H=125.7$ GeV, we finally find for this sum: 
$2.46$.

As for the first two terms in the second line of the 
Eq.\,(\ref{sumsum}), they are nothing but the 
quadratic and the logarithmic divergences respectively, 
and can  be computed with the help of ordinary momentum integrals
(Fourier space). By computing these integrals within 
the framework of the $\overline{MS}$ scheme, and determining the 
counterterms accordingly, we have

\begin{eqnarray}
&&\frac{1}{2}\mbox{Tr}\left[(-\partial^2)^{-1}
V''(\phi_b)\right]
-\frac{1}{4}\mbox{Tr}\left[(-\partial^2)^{-1}
V''(\phi_b)(-\partial^2)^{-1}V''(\phi_b)\right]-
\delta S_H^{ct,\overline{MS}}\nonumber\\
&=&[(1+L)I_1+I_2]\,,
\end{eqnarray}
where $L=\ln\left(\mu R_{M}e^{\gamma_E}/2\right)$, 
$\gamma_E$ is the Euler gamma and 
\begin{eqnarray}\label{Iintegral}
I_1&=&\frac{1}{32}\int\frac{d^4q}{(2\pi)^4} 
{\tilde {V}''} (-q)\tilde V''(q)\nonumber\\
I_2&=&\frac{1}{32}\int\frac{d^4q}{(2\pi)^4} 
\tilde V'' (-q)\tilde V''(q)
\,\ln\left(\frac{2 e^{-\gamma_E}}{(q^2)^{1/2}R_M}\right).
\end{eqnarray}
where $\tilde V''(q)$  is the Fourier transform of 
$V''(\phi_b(r))$.
For the potential in Eq.\,(\ref{potentialvecchio}), i.e.
for the potential of the SM alone, the integrals in 
Eq.\,(\ref{Iintegral}) can be computed analytically and
we find $I_1=-3$ and $I_2=1/2$. 
The renormalized sum of Eq.\,(\ref{sumrenorm}) is then 
given by
\bea \label{resumHSM}
\left[\frac{1}{2}\sum_{j=0}^{\infty} 
(2j+1)^2\ln \rho_j\right]^{SM}_{r}&=&6.02-\frac{5}{2}-3 L\,.
\eea

Putting together then the results of Eq.\,(\ref{resumHSM}),
with those of Eqs.\,(\ref{jtransnumSM}), (\ref{jdilnumSM}) 
and (\ref{jsu2numSM}),
and choosing the renormalization scale (as mentioned above)
so to make the logarithmic term vanishing ($L=0$), we finally 
get 
\begin{equation}
\Delta S_H^{SM}=-5.88792\,.
\end{equation}

For the potential with new physics terms, 
Eq.\,(\ref{potentialnuovo}), on the contrary,  
both $I_1$ and $I_2$ have to 
be computed by means of some numerical routine, and the result
depends on the value of the couplings. 
For the value of the parameters given in the 
text ($\lambda_6=-2$ and $\lambda_8=2.1$), we get:
 $I_1=-6.19$ and $I_2=8.92$.
The renormalized sum in  Eq.\,(\ref{sumrenorm}) is now
given by
\be \label{resumHnew}
\left[\frac{1}{2}\sum_{j=0}^{\infty} 
(2j+1)^2\ln \rho_j\right]^{new}_{r}=2.72856 - 6.19251\cdot L
\ee

For the purpose of comparing the two results (with and without the 
new physics operators), we choose even for this case the same 
renormalization scale taken above, namely $\mu_{ren} =
2\, e^{-\gamma_E}/ R^{SM}_M\simeq 2 \times 10^{17} GeV$. The 
logarithmic term $L$ in this case is not vanishing, 
as $R^{new}_{M}$ is different from $R^{SM}_M$.  
Putting together then the result of Eq.\,(\ref{resumHnew})
with those of and of Eqs.\,(\ref{jtransnumnew}), 
(\ref{jsu2numnew}), we finally get ($L=-2.63$) 
\begin{equation}
\Delta S_H^{new}=-9.4425\,.
\end{equation}

For the evaluation of $\Delta S_t$ and $\Delta S_{gg}$ 
in Eqs.\,(\ref{deltast}) and (\ref{deltasgg}), we have 
to follow steps very similar to those used for  
$\Delta S_{H}$. The only novelty is that 
we now have to deal also with (Dirac and/or Lorentz) 
indices, the eigenfunctions of the corresponding 
fluctuation operators, $S''_t[\phi_b]$ and $S''_{gg}[\phi_b]$, 
having an additional algebraic, spinor or vector, 
structure that can de dealt with in a standard 
manner\,\cite{dirtens}. 

When we consider the $SM$ theory only (SM couplings only), i.e. 
when  the potential of the scalar sector is given by 
Eq.\, (\ref{potentialnuovo}),
the expression for the renormalized determinant appearing 
in $\Delta S_t$ only depends on the ratio of the top Yukawa
coupling to the quartic coupling, 
${g_t^2}/{|\lambda|}$,  and turns out to be

\bea\label{determSMt}
\left[-\frac{3}{2}\ln\left(\frac{\mbox{Det}S''_t[\phi_b]}{\mbox{Det}S''_t[0]}\right)\right]^{SM}_r
=F_t\left(\frac{g^2_t}{|\lambda|}\right)+\frac{g^4_t}{\lambda^2}
\left(\frac{5}{6}+L\right)+\frac{g^2_t}{|\lambda|}
\left(\frac{13}{6}+2L\right)\,,
\eea
where $F_t$ is a numerical function.
For the central experimental values of $M_H$ and $M_t$,
$M_H=125.7$ GeV and $M_t=173.34$ GeV, 
we find
that $g_t$ at the scale 
$\mu_{ren} =2 e^{-\gamma_E}/ R_M\simeq 2 \times 10^{17} GeV$
is  $g_t=0.40375$ and that 
$g_t^2/|\lambda|\simeq 12.1184$, and the corresponding  
$F_t$ is $F_t(g^2_t/|\lambda|)\simeq -193.058$.
From Eq.\,(\ref{determSMt}) then, $\Delta S_t$ when only 
SM opertors are considered turns out to be
\be
\Delta S_t^{SM} \simeq  -19.29\,. 
\ee

When we consider the potential that involves the 
contribution of new physics operators, i.e. the potential 
of equation (\ref{potentialnuovo}) that contains the
contribution of $\lambda_6$ and the $\lambda_8$, $\Delta S_t$
has to be computed in a way that is similar to the one 
used for the Higgs sector, i.e. for $\Delta S_H$. We find
\be
\Delta S_t^{new}\simeq -4.98315\,.
\ee

Finally, we have to consider $\Delta S_{gg}$. When the SM 
interactions only are taken into account, 
the renormalized determinant appearing in $\Delta S_{gg}$ 
turns out to depend on the two ratios $\frac{g^2}{|\lambda|}$ 
and $(g^2+g^{'2})/|\lambda|$, and we have 
\bea \label{determSMgg}
&&\left[\frac{1}{2}\ln\left(\frac{1}{R_M^{6}}\frac{\mbox{Det}'S''_{gg}[\phi_b]}{\mbox{Det}S''_{gg}[0]}\right)-\frac{1}{2}\ln\left(\frac{\mbox{Det}S''_{ghost}[\phi_b]}{\mbox{Det}S''_{ghost}[0]}\right)\right]^{SM}_r\nonumber\\
&=&\left\{F_g(g^2/|\lambda|) -\left(\frac{6L+5}{9}
+\frac{7+6L}{9}\frac{g^2}{|\lambda|}+\frac{1+2L}{16}\frac{g^4}{\lambda^2}\right)\right\}\nonumber\\
&+&\frac{1}{2} \times \left\{\frac{g^2}{|\lambda|}\rightarrow
\frac{g^2+g^{'2}}{|\lambda|}\right\}
\eea
where again $F_g$ is a numerical function.
We find that the renormalized couplings at the renormalization 
scale $\mu_{ren} =2 e^{-\gamma_E}/ R_M\simeq 2 
\times 10^{17}$ GeV are $g=0.5168$ and $g'=0.459068$, 
that in turn gives $g^2/|\lambda|\simeq 19.8562$  and 
$(g^2+g^{'2})/|\lambda|\simeq 35.5228$. 
Moreover,  $F_g(g^2/|\lambda|)\simeq 93.9308$ and  
$F_g((g^2+g^{'2})/|\lambda|)\simeq 380.344$.
Therefore, putting together these results with those 
of Eq.\,(\ref{jsu2numSM}) we find
\be
\Delta S^{SM}_{gg}\simeq 67.4064\,.
\ee

Once again, when we consider the potential (\ref{potentialnuovo}) 
with the contribution of new physics interactions, 
and therefore the contribution of the additional couplings
$ \lambda_6$ and the $ \lambda_8$, the expression corresponding 
Eq.\,(\ref{determSMgg}) can be computed only numerically. Performing
this computation, and then including the contribution of 
Eq.\,(\ref{jsu2numnew}), we finally find
\be
\Delta S_{gg}^{new}\simeq 8.42902\,.
\ee

This latter result completes the work of this Appendix.
Actually, by collecting all of the quantum 
fluctuation contributions $\Delta S_i$, discussed in the 
present appendix, the tables for the loop contribution to
$\tau$ presented in section II and III are obtained.  

\vskip 20pt
\section{} \label{App3}
\vskip 15pt

In this appendix we present the numerical
determination of the bounce solution to Eq.\,(\ref{eqcom}) 
of Section III in the text, with boundary conditions given 
by Eqs.\,(\ref{c1}) and (\ref{c2}).   
These boundary 
conditions at $x=0$ and $x=\infty$ are implemented by first 
considering a minimal and a maximal value of $x$, $x_{min}$ and 
$x_{max}$, and then studying the convergence of the solution 
(to the desired level of accuracy) by taking lower and lower 
values of $x_{min}$ and higher and higher values of $x_{max}$.    
As described in Ref.\,\cite{sherrep}, one technique is to 
guess values of $\phi(0)$ and integrate outward. If the value 
of $\phi(0)$ is too large, then $\phi$ will overshoot the 
value of $\phi$ at the false vacuum, whereas if it is too small, 
it will undershoot. So one can gradually converge on the 
correct value. However, the forward-backward shooting 
method converges more quickly.
 
To proceed with such an analysis, however, we first need 
to study analytically the asymptotical behavior of Eq.\,(\ref{eqcom}) around
$x=0$ and $x=\infty$. Let us begin by performing an expansion of 
$\varphi(x)$ in powers of $x$ around $x=0$. For our purposes, it is
sufficient to consider 
an expansion up to $x^8$. We write only the first few terms, 
\begin{eqnarray}
\varphi(x)=B_0 + B_2 x^2 + B_3 x^3+ \cdots \label{series}
\end{eqnarray}
where, due to the condition $\varphi'(0)=0$, the linear 
term is missing. Inserting the expansion \,(\ref{series}) in 
(\ref{eqcom}), we find that the coefficients of odd 
powers of $x$ vanish, while those of even powers of $x$ are 
all given in terms of $B_0$ (from now on indicated with $B$): 
\begin{eqnarray}
\varphi(x)=B+\left(\lambda B^3+\lambda_6 B^5+
\lambda_8 B^7\right)\frac{x^2}{8} +\dots\,, \label{lowbound}
\end{eqnarray}
where only the first and the second term of the expansion are 
explicitly written. 

As we shall see in a moment, the coefficient of $x^2$ (for the 
case of interest to us) is negative and Eq.\,(\ref{lowbound}) shows
that, for values of $x$ close to $x=0$,  the bounce behaves as 
an upside down parabola. This observation is very useful for our 
numerical analysis.

Let us study now the asymptotic region $x \to \infty$. As the bounce 
has to fulfill the condition (\ref{c1}), we expand $\varphi(x)$ in 
powers of $1/x$. For our purposes, we perform the expansion up 
to $1/x^{20}$. Writing again only the first few terms,  
\begin{eqnarray}
\varphi(x)&=& \frac{A_1}{x}+ \frac{A_2}{x^2}+ 
\frac{A_3}{x^3}+\frac{A_4}{x^4}+\dots
\label{largesvil}
\end{eqnarray}

Inserting the expansion \,(\ref{largesvil}) in (\ref{eqcom}), 
we find that the coefficients of odd powers of $1/x$ vanish, while
those of even powers are all written in terms of $A_2$ (from now on
indicated with $A$)
\begin{eqnarray}
\varphi(x)=\frac{A}{x^2}-\frac{\lambda}{8}
\frac{A^3}{x^4}+\dots\label{upbound},
\end{eqnarray}
where, as for Eq.\,(\ref{lowbound}), only the first and the second 
term are explicitly written. 
Eq.\,(\ref{upbound}) shows that, for large values of $x$,
$\varphi(x)$ behaves as
$1/x^2$. As we shall see in a moment, this observation is 
very useful for our numerical analysis.

Let us proceed now with the forward-backward shooting. 
Going back to 
Eq.\,(\ref{lowbound}), we choose a value of $x$ close to $x=0$,
say $x=x_{min} <<1$, and consider the two
``initial conditions'' $\varphi(x_{min})$  and $\varphi'(x_{min})$
\begin{eqnarray}
\varphi(x_{min})&=&B+\left(\lambda B^3+\lambda_6 B^5+
\lambda_8 B^7\right)\frac{x_{\min }^2}{8} +\dots \nonumber\\
\varphi'(x_{min})&=&\left(\lambda B^3 
+ \lambda_6 B^5 +\lambda_8 B^7\right)\frac{x_{\min }}{4}+\dots 
\label{lowboundmm}
\end{eqnarray}
for the integration of the second order differential equation 
(\ref{eqcom}). Choosing also a value $x=x_{max} >>1$,  
Eq.\,(\ref{eqcom}) is integrated, for different choices of $B$,
in the range $[x_{min}$, $x_{max}]$. 

\begin{figure}[htp]
\includegraphics[width=0.75\textwidth,angle=0]{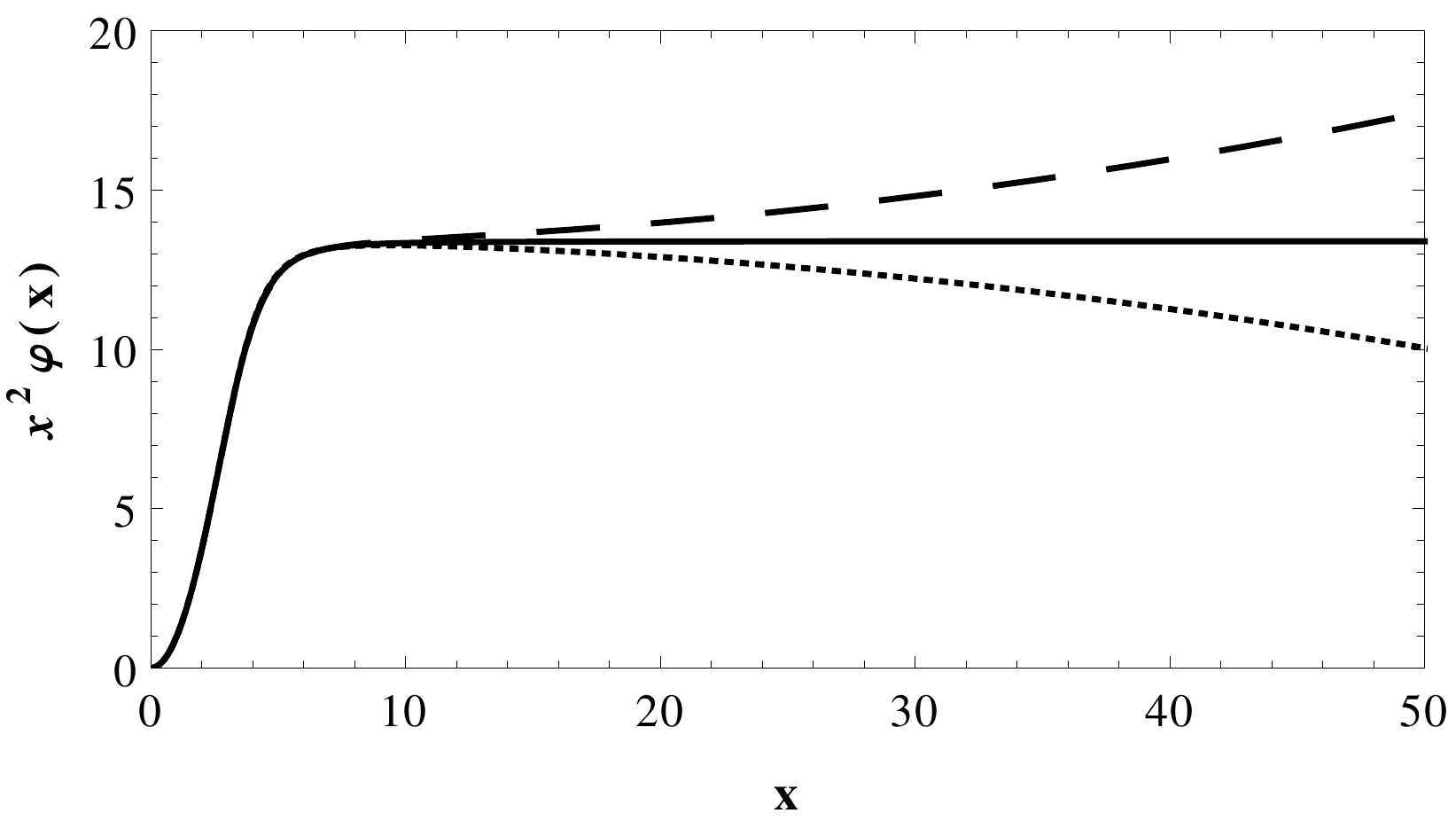}
\caption{Plot of $x^2 \varphi(x)$, for three different solutions 
of Eq.\,(\ref{eqcom}),
with $\lambda=-0.01345$, $\lambda_6=-2$ and $\lambda_8=2.1$.
The $x$ range goes from 
$x=x_{min}= 6 \times 10^{-2}$ to $x=50$, although the 
numerical integration is performed up to 
$x_{max}= 10^{2}$. 
This figure well illustrates the forward shooting.
Eq.\,(\ref{eqcom}) is integrated starting with the initial values 
\,(\ref{lowboundmm}) for $\varphi(x_{min})$ and $\varphi'(x_{min})$
at $x=x_{min}= 6 \times 10^{-2}$. The parameter $B$ is tuned 
until $x^2 \varphi(x)$ saturates to a plateau for values of 
$x$ greater than $x_{min}$ and at least up to  
$x_{max}$. We see that for $B=0.967$ (dotted line) and
$B=0.9665$ (dashed line), $x^2 \varphi(x)$ diverges downwards and 
upwards, respectively. For $B=0.966777$ (solid line), the plateau 
is reached and our first approximation to the bounce is obtained
.}\label{gooshootinga}
\end{figure}

As from (\ref{upbound}) 
we know that, for large values of $x$, $\varphi(x)$ behaves 
as $1/x^2$, the search for the bounce is realized by 
{\it tuning} $B$ so that, 
for large values of $x$ (actually up to $x_{max}$), the 
product $x^2 \varphi(x)$ reaches a plateau. This completes
the ``forward'' part of the method. For the ``backward'' part, we 
have to follow similar steps, but starting from large values 
of $x$ and integrating back our differential equation (\ref{eqcom})
towards small values.

Let us study now this equation for the values of the coupling 
constants considered in the text, namely $\lambda=-0.01345$, 
$\lambda_6=-2$ and $\lambda_8=2.1$. 
The forward shooting described above is illustrated in 
Fig.\,\ref{gooshootinga}, where $x^2 \varphi(x)$ is plotted 
against $x$. For the integration range, we have chosen 
$x_{min}= 6 \times 10^{-2}$, \, $x_{max}=10^{2}$. 

The central part of the forward shooting is the tuning of 
the parameter $B$. In Fig.\,\ref{gooshootinga}, we plot three 
curves $x^2 \varphi(x)$ for three different values of $B$. Although 
the $x$ range in the figure goes from $x=x_{min}= 6 
\times 10^{-2}$ to $x=50$, the numerical integration is performed 
from $x_{min}=6\times 10^{-2}$ up to $x_{max}= 10^{2}$. 
The dotted line is obtained for $B=0.967$. After a first transient 
regime, from $x=x_{min}$ up to $x\sim 5$, 
the product $x^2\varphi(x)$ becomes almost constant in the range 
from $x\sim 5$ to $x\sim 10$. For $x>10$, however, it starts to 
decrease, so that the corresponding $\varphi(x)$ does not satisfy 
the asymptotic condition $\varphi(x)\propto 1/x^2$.

For a lower value of $B$, $B=0.9665$, the product 
$x^2\varphi(x)$ is given by the dashed line of 
Fig.\ref{gooshootinga}. Again, after a first transient regime, 
$x^2\varphi(x)$ becomes almost constant in the range  
from $x\sim 5$ to $x\sim 10$. For $x>10$, however, $x^2\varphi(x)$ 
starts to increase, again violating the asymptotic condition 
$\varphi(x)\propto 1/x^2$.
Finally, continuing with the tuning of $B$, it is found that, 
for $B=0.966777$ (solid line), the product $x^2\varphi(x)$, 
turns out to reach a plateau up to $x=x_{max}$ (in the figure 
the $x$ range is extended only up to $x=50$).  
The corresponding numerical solution $\varphi(x)$  
is then our first estimate of the bounce (in the 
range $x_{min}\leq x\leq x_{max}$).

\begin{figure}[htp]
\includegraphics[width=0.75\textwidth,angle=0]{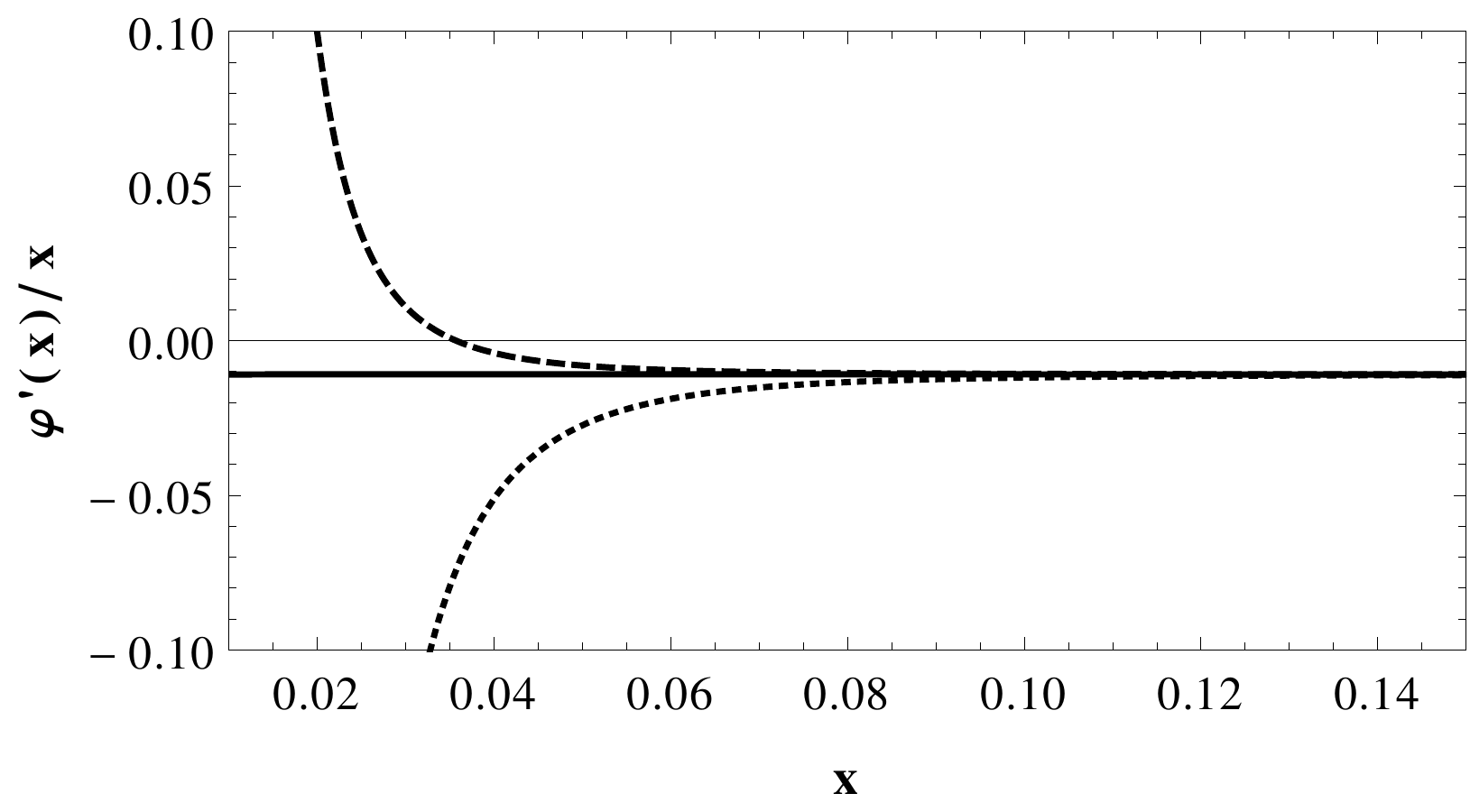}
\caption{This figure illustrates the backward shooting with a 
plot of $\varphi'(x)/x$  for three different solutions of 
Eq.\,(\ref{eqcom}) ($\lambda=-0.01345$, $\lambda_6=-2$, 
$\lambda_8=2.1$).
The $x$ range goes from $x_{min}= 10^{-2}$ to 
$x=0.15$, although the 
numerical integration is performed from $x_{max}=10^{2}$
down to $x_{min}= 10^{-3}$. 
Eq.\,(\ref{eqcom}) is integrated with initial values 
\,(\ref{upboundmm}) for $\varphi(x_{max})$ and $\varphi'(x_{max})$. 
The parameter $A$ is tuned 
until $\varphi'(x)/x$ saturates to a plateau for small values of $x$. 
For $A=13.39776497$ (dotted line) and $A=13.39776498$ (dashed line), 
$\varphi'(x)/x$ diverges downwards and upwards, respectively. 
Finally, for $A=13.3977649785377$ (solid line), the plateau 
is reached. We have then, to a very high degree of 
numerical accuracy, the bounce solution to our equation.}
\label{gooshootingb}
\end{figure}

The next step of our numerical procedure is the backward shooting, 
where we integrate backward Eq.\,(\ref{eqcom}) 
from the upper limit $x_{max}$ of the 
previous (forward) integration, $x_{max}= 10^2$, and 
extend the integration domain down to 
$x'_{min}= 10^{-3} < x_{min}$. 
The initial conditions are taken from 
the asymptotic behavior of the bounce, Eq.\,(\ref{upbound}),
\begin{eqnarray}
\varphi(x_{max})&=&\frac{A}{x_{\max }^2}-\frac{\lambda}{8}
\frac{A^3}{x_{\max }^4}+\dots\nonumber\\
\varphi'(x_{max})&=&-\frac{2 A}{x_{\max }^3}+\frac{\lambda}{2}
\frac{A^3}{x_{\max }^5}+\dots\,. \label{upboundmm}
\end{eqnarray}

Similarly to the forward case, we have to fine tune the 
parameter $A$ so that, according to (\ref{lowbound}), 
the solution $\varphi(x)$, for small values of $x$, 
satisfies the condition 
\begin{eqnarray}\label{valueA}
\frac{\varphi'(x)}{x} \simeq Const.
\end{eqnarray} 
in the range $[x'_{min},x_{max}]$.

In Fig.\,\ref{gooshootingb} we plot $\varphi'(x)/x$ 
versus $x$ for three different values of $A$ and  
illustrate how the fine tuning of $A$ is realized.
The domain of our numerical (backward) integration ranges 
from $x_{max}= 10^{2}$ down to $x'_{min}= 10^{-3}$,
although in the figure we only show the range from 
$x'_{min}= 10^{-3}$ to $x=0.15$.

The dotted line is obtained for $A=13.39776497$. As we approach 
smaller and smaller values of $x$, $\varphi'(x)/x$ starts to 
decrease, thus violating the bounce condition 
$\varphi'(x)/x \sim Const.$. The dashed line is obtained for 
$A=13.39776498$. For smaller and smaller values of $x$, 
$\varphi'(x)/x$ starts to increase, again violating the bounce 
condition. Finally, for $A=13.3977649785377$, the ratio 
$\varphi'(x)/x$ reaches a plateau, thus showing that this 
is the value of $A$ that corresponds to the bounce solution
(at this order of numerical precision). 

We can then iterate the procedure of forward and backward 
integrations by enlarging the range of integration, thus 
obtaining values of $A$ and $B$ with higher and higher degree 
of numerical accuracy.

\vskip 20pt
\section{} \label{App4}
\vskip 15pt

Here we consider a toy grand unified model which gives Eq.\,(\ref{potnew}) as the effective low energy theory.    Note that nothing we have done in this paper involves gravity, and thus $M_P$ can be replaced by the unification scale, $M_X$.   Note that if $M_X << M_P$, the effective values of $\lambda_6$ and $\lambda_8$ would be much larger,  leading to even bigger effects, and thus the conservative approach is to consider the case in which $M_X \sim M_P$.

We will consider the minimal $SU(5)$ model broken at the $M_P$ scale.   Such a model, of course, is phenomenologically unacceptable, but if this model gives the potential of Eq.\,(\ref{potnew}) with $O(1)$ coefficients, then clearly a more complicated (and acceptable) grand unified theory can also do so.   The symmetry is broken down to $SU(3)\times SU(2)\times U(1)$ with the minimal Higgs content of a 24-plet, and the breaking of the Standard Model group uses a 5-plet.

The Higgs potential is given, with $\Psi$ being the 24 and $\phi$ being the 5, by
\begin{equation}
V(\Psi) = -\frac{1}{2} \mu^2 {\rm Tr}(\Psi^2) + \frac{1}{4}a({\rm Tr}(\Psi^2))^2 + \frac{1}{2} b{\rm Tr}(\Psi^4)
\end{equation}
\begin{equation}
V(\phi) = -\frac{1}{2} \nu^2 \phi^\dagger \phi + \frac{1}{4} \lambda (\phi^\dagger \phi)^2
\end{equation}
\begin{equation}
V(\Phi,\phi) = \alpha \phi^\dagger \phi {\rm Tr} (\Psi^2) + \beta \phi^\dagger \Psi^2 \phi
\end{equation}
The relevant Higgs fields in the 24 are the $\Psi_3$ and the $\Psi_0$, where $\Psi_3$ is the neutral member of the color-singlet, isotriplet and $\Psi_0$ is the isosinglet.

\begin{figure}
${}$\vskip1cm \epsfig{file=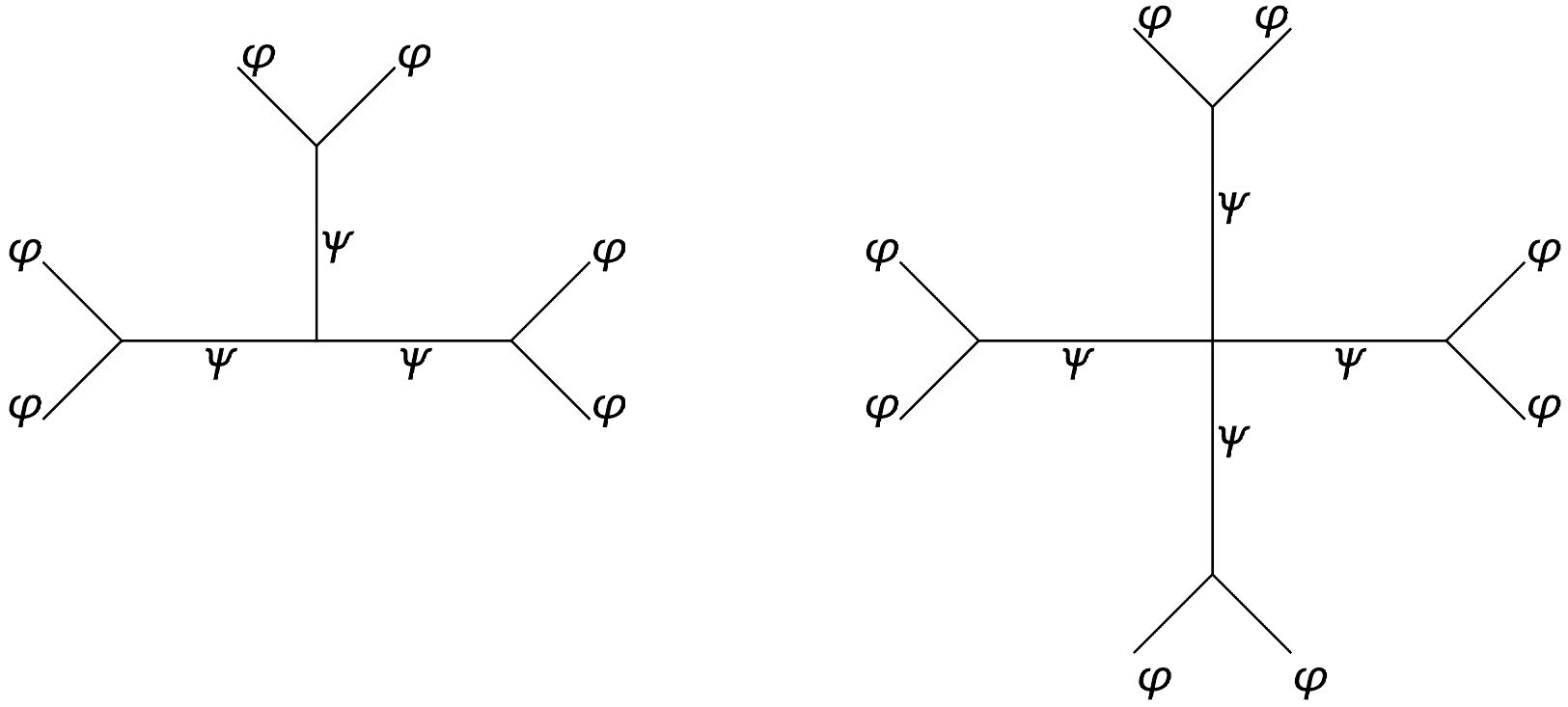,width=.75\hsize}
\caption{Diagrams leading to higher dimensional operators in the low energy theory.  $\phi$ is the Standard Model Higgs and $\Psi$ is the 24-plet.}\label{lines}
\end{figure}

The diagrams leading to higher order operators in the effective low-energy theory (below $M_P$) to leading order in the couplings are shown in Fig. \ref{lines}.   For the $\phi^6$ term, there are two diagrams, one with three $\Psi_0$ fields and one with two $\Psi_3$ fields and one $\Psi_0$ field.    Using the vertices found in Ref.  \cite{Huffel:1981ux}, we find that the contributions to $\lambda_6$ are
\begin{equation}
5\frac{(\frac{1}{4}\alpha+\frac{3}{40}\beta)^3}{(15a+7b)^2}
\end{equation} for the first, and
\begin{equation}
\frac{1}{10}\left(\frac{3}{4}\right)^4\frac{5a+9b}{15a+7b}(\frac{1}{10}\alpha+\frac{3}{100} \beta)\frac{\beta^2}{b^2}
\end{equation} for the second.
We have chosen the scale $M_P$ to equal the vev of the 24-plet (which is numerically very close to the gauge boson mass).

Now, in order to have the correct symmetry breaking pattern, $\beta$ must be negative, and $15a+7b$ and $b$  must be positive.   But $\alpha+\frac{3}{10}\beta$ can have either sign.    So if $\alpha$, for example, equals $\pm 4$ (well below the unitarity bound, see \cite{Huffel:1981ux}), $\beta$ is small, and $15a+7b$ is, say, 1, then the contribution to $\lambda_6$ is 
$\pm 2$, 
showing that a large coefficient isn't unreasonable, and well within unitarity limits.   Of course, the contribution to $\lambda_6$ would be even larger if, as expected, the unification scale is well below the Planck scale.

For the $\phi^8$ term, one has three diagrams, one with four $\Psi_0$, one with four $\Psi_3$ and one with two of each  (there are six copies from combinatorics).  The contributions to $\lambda_8$ are
\begin{equation}
{\frac87}
\frac{(\frac{1}{4}\alpha + \frac{3}{40}\beta)^4}{(15a+7b)^3}
\end{equation} from the first.  This term numerically dominates for most of parameter-space.
The second gives
\begin{equation}
\frac{\left(\frac{3}{40}\beta\right)^4}{(2b)^4}
\end{equation}
and the third gives
\begin{equation}
\left(\frac{3}{20}\right)^4\frac{(5a+9b)(\frac{1}{4}\alpha+\frac{3}{40}\beta)^2}{(10b)^2(15a+7b)^2}
\end{equation}

Again, these can easily be large and still be within unitarity bounds, even if the unification scale is at the Planck scale.   Note that the expressions are positive, and thus Eq.\,(\ref{potnew}) would be bounded.   Also note that, to leading order, there are no $\phi^{10}$ terms, further justifying the truncation in Eq.\,(\ref{potnew}).   

This model is not to be taken too seriously, of course, but does demonstrate how a very simple unified theory can give the effective low energy theory of Eq.\,(\ref{potnew}).


\begin{thebibliography}{100}

\bibitem{cabibbo} N. Cabibbo, L. Maiani, G. Parisi, R. Petronzio, 
Nucl.Phys. B158 (1979) 295.
\bibitem{flores} R. A. Flores, M. Sher, Phys. Rev. D27 (1983) 1679.
\bibitem{lindner} M. Lindner, Z. Phys. 31 (1986) 295.
\bibitem{bennet} D.L. Bennett, H.B. Nielsen and I. Picek, 
Phys. Lett. B 208 (1988) 275.
\bibitem{sherrep} M. Sher, Phys. Rep. 179 (1989) 273.
\bibitem{lindsher} M. Lindner, M. Sher, H. W. Zaglauer, Phys. Lett. B228 (1989) 139.
\bibitem{arnold} P. B. Arnold, Phys. Rev. D {40} (1989) 613.
\bibitem{anderson} G. Anderson, Phys. Lett. B243 (1990) 265.
\bibitem{arnoldvok} P. Arnold and S. Vokos, Phys. Rev. D44 (1991) 3620.
\bibitem{ford} C. Ford, D.R.T. Jones, P.W. Stephenson, M.B. Einhorn, Nucl.Phys. B395 (1993) 17.
\bibitem{sher2} M. Sher, Phys. Lett. B317 (1993) 159.
\bibitem{altar} G. Altarelli, G. Isidori, Phys. Lett. B337 (1994) 141.
\bibitem{casas1} J.A. Casas, J.R. Espinosa, M. Quir\'os, Phys. Lett. B342. (1995) 171.
\bibitem{espiquiros} J.R. Espinosa, M. Quir\'os, Phys.Lett. B353 (1995) 257.
\bibitem{casas2} J.A. Casas, J.R. Espinosa, M. Quir\'os, 
Phys. Lett. B382. (1996) 374.
\bibitem{frogniel1} C. D. Froggatt and H. B. Nielsen, 
Phys. Lett. B 368 (1996) 96.
\bibitem{frogniel2} C.D. Froggatt, H. B. Nielsen, Y. Takanishi, 
Phys.Rev. D64 (2001) 113014.
\bibitem{isido} G. Isidori, G. Ridolfi, A. Strumia, 
Nucl. Phys. B609 (2001) 387.
\bibitem{espigiu} J.~R.~Espinosa, G.~F.~Giudice and A.~Riotto,
JCAP  0805 (2008) 002.
\bibitem{ellisespi}
J.~Ellis, J.~R.~Espinosa, G.~F.~Giudice, A.~Hoecker and A.~Riotto,
Phys.\ Lett.\ B  679 (2009) 369.
\bibitem{isiuno} J. Elias-Miro, J.R. Espinosa, G.F. Giudice, G. 
Isidori, A. Riotto, A. Strumia, Phys. Lett. B709 (2012) 222. 
\bibitem{isidue} G. Degrassi, S. Di Vita, J. Elias-Miro, 
J. R. Espinosa, G. F. Giudice, G. Isidori, A. Strumia, JHEP 1208 (2012) 098.
\bibitem{degrassi} D. Buttazzo, G. Degrassi, P. P. Giardino, 
G. F. Giudice, F. Sala, A. Salvio, A. Strumia, JHEP 1312 (2013) 089.
\bibitem{atlas} ATLAS Collaboration, Phys. Lett. B710 (2012) 49.
\bibitem{cms} CMS Collaboration, Phys. Lett. B710 (2012) 26. 
\bibitem{abdel}  S.~Alekhin, A.~Djouadi and S.~Moch,
  Phys.\ Lett.\ B  716 (2012) 214.
\bibitem{degrassi2} G.~Degrassi,
  arXiv:1405.6852 [hep-ph].
\bibitem{ber} F.L. Bezrukov, M. Shaposhnikov, Phys.Lett. B659 (2008) 703; 
JHEP 0907 (2009) 089; F.L. Bezrukov, A. Magnin, M. Shaposhnikov, 
Phys.Lett. B675 (2009) 88.
\bibitem{cole1} S. Coleman, Phys. Rev. D15 (1977) 2929. 
\bibitem{Frampton:1976kf}
  P.~H.~Frampton, Phys.\ Rev.\ Lett.\   37 (1976) 1378 [Erratum-ibid.
\  37 (1976) 1716].
\bibitem{cole2} C. Callan, S. Coleman, Phys. Rev. D16 (1977) 1762.
\bibitem{brames} V. Branchina, E. Messina, Phys.Rev.Lett. 111 (2013) 241801. 
\bibitem{bra} V.~Branchina,  arXiv:1405.7864 [hep-ph].
\bibitem{bramespla} V. Branchina, E. Messina, A. Platania, 	
arXiv:1407.4112 [hep-ph].

\bibitem{Kusenko:1996bv} A.~Kusenko, K.~M.~Lee and E.~J.~Weinberg,
Phys.\ Rev.\ D  55 (1997) 4903 [hep-th/9609100].
\bibitem{miha} L.N. Mihaila, J. Salomon and M. Steinhauser, 
Phys. Rev. Lett. 108 (2012) 151602. 
\bibitem{chety} K. Chetyrkin and M. Zoller, JHEP 06 (2012) 033.
\bibitem{shapo} F. Bezrukov, M. Yu. Kalmykov, B. A. Kniehl, 
M. Shaposhnikov, JHEP 1210 (2012) 140.
\bibitem{dunne} G.V. Dunne, H. Min, Phys. Rev. D72, (2005) 125004.
\bibitem{wein} K. Lee, E.J. Weinberg, Nucl. Phys. B267 (1986) 181.
\bibitem{yaglom} G. Dunne, J. Phys. A: Math. Theor. 41 (2008) 304006.
\bibitem{dirtens} A. Pais, Proc. Natl. Acad. Sci., 40 (1954) 835; 
M. Daumens, P. Minnaert, Jour. Math. Phys., 17 (1976) 1903.
\bibitem{Huffel:1981ux} H.~Huffel, 
Z.\ Phys.\ C  10 (1981) 327.

\end{thebibliography}
\end{document}